\documentclass[aps,prl,twocolumn,superscriptaddress,groupedaddress]{revtex4}  
\usepackage{graphicx}  
\usepackage{dcolumn}   
\usepackage{bm}        
\usepackage{amssymb}   
\usepackage{amsmath}
\usepackage{upgreek}
\begin{document}


\title{Direct Mapping of Local Seebeck Coefficient in 2D Material Nanostructures via Scanning Thermal Gate Microscopy}

\author{Achim~Harzheim}
\affiliation{Department of Materials, University of Oxford, Oxford OX1 3PH, United Kingdom}
\author{Charalambos Evangeli}
\affiliation{Department of Materials, University of Oxford, Oxford OX1 3PH, United Kingdom}
\affiliation{Department of Physics, Lancaster University, Bailrigg, LA1 4YB, Lancaster, United Kingdom}
\author{Oleg V. Kolosov}
\email{o.kolosov@lancaster.ac.uk}
\affiliation{Department of Physics, Lancaster University, Bailrigg, LA1 4YB, Lancaster, United Kingdom}
\author{Pascal~Gehring}
\email{pascal.gehring@imec.be}
\affiliation{Department of Materials, University of Oxford, Oxford OX1 3PH, United Kingdom}
\affiliation{Kavli Institute of Nanoscience, Delft University of Technology, Delft 2628, Netherlands}
\affiliation{Present address: IMEC, Kapeldreef 75, 3001 Leuven, Belgium}


\begin{abstract}
\noindent Local variations in the Seebeck coeﬃcient in low-dimensional materials-based nanostructures and devices play a major role in their thermoelectric performance. Unfortunately, currently most thermoelectric measurements probe the aggregate characteristics of the device as a whole, failing to observe the effects of the local variations and internal structure. Such variations can be caused by local defects, geometry, electrical contacts or interfaces and often substantially influence thermoelectric properties, most profoundly in two-dimensional (2D) materials. Here, we use Scanning Thermal Gate Microscopy (STGM), a non-invasive method not requiring an electrical contact between the nanoscale tip and the probed sample, to obtain nanoscale resolution 2D maps of the thermovoltage in graphene samples. We investigate a junction formed between single-layer and bilayer graphene and identify the impact of internal strain and Fermi level pinning by the contacts using a deconvolution method to directly map the local Seebeck coefficient. The new approach paves the way for an in-depth understanding of thermoelectric behaviour and phenomena in 2D materials nanostructures and devices.
\end{abstract}

\maketitle



%
\noindent Increasing the efficiency of thermoelectric materials is necessary to enable devices that could scavenge waste heat and convert it into useful electrical energy. This is particularly true as current bulk thermoelectric generators lack the required conversion capability to be competitive with everyday heat engines such as standard combustion engines \cite{Zhang2014}.$\,\,$In addition, knowledge about the thermoelectric properties of materials and devices can reveal fundamental physical characteristics \cite{Crossno2016} or be used to design novel temperature sensors \cite{Harzheim2020_1} and increase device efficiency in e.g. photovoltaic/thermoelectric hybrid structures \cite{Kraemer2008}.$\,\,$While device-scale summative thermoelectric measurements, such as measuring the voltage drop over a sample in response to an applied temperature gradient, are common, they cannot reveal the local variations of the thermoelectric properties in the sample, that in two-dimensional (2D) materials can vary on the length scale of few tens of nanometres.In particular, it has been shown that metal contacts \cite{Xia2009}, defects and other localized effects \cite{Guo2018,Levander2011} can impact the local thermoelectric characteristics drastically. \\
Thus, in order to study thermoelectric phenomena in 2D materials nanostructures, it is of ultimate importance to study them locally, with currently very limited choices of approaches available for such a measurement. Whereas some SThM techniques to measure local thermoelectric phenomena have been reported previously, these require a direct electrical contact between a conducting SThM tip and the sample thus adding a major variability to thermoelectric measurements due to the unstable nature of scanning nanoscale electrical contacts especially to non-metallic surfaces \cite{Lyeo2004,Zhang2010}. Another attempt at the local measurement of thermoelectric effects was made via photocurrent measurements. However, laser spot sizes tend to be in the $\upmu$m range, the temperature increase caused by the laser can typically only be simulated with finite element analysis and the technique is inherently measuring a mix of photo thermoelectric and photovoltaic effects \cite{Sun2012,Gabor2011}.\\
The ideal local measurement of thermoelectric phenomena in 2D materials nanostructures would require only a thermal contact between the probe and the surface of the device, provide a spatial resolution on the order of few or few tens of nanometres, and necessitate only low contact forces to avoid damaging the sample surface. The intrinsic atomically smooth nature of 2D materials will then provide a natural test bed for such localized thermoelectric measurements. In addition, 2D materials allow to readily tune their electronic properties via a back-gate, have an inherent control over their layer thickness and exhibit an ordered crystalline structure. 2D materials are also particularly promising as they have the potential to be integrated on a wafer scale basis \cite{Zhang2014_1, Avsar2011} or to be incorporated into current CMOS technology \cite{Goossens2017}. Graphene is the most-studied material out of the 2D family, and its thermoelectric properties can be influenced by gating \cite{Zuev2009,Wei2009}, local charge carrier fluctuations \cite{Woessner2016}, introduction of nanoparticles \cite{Shiau2019} or increased scattering at the edges \cite{Harzheim2018}. Yet, the local thermoelectric properties of graphene have mostly been studied globally or in combination with photocurrent measurements.
\begin{figure}
	\centering
	\includegraphics[width=\linewidth]{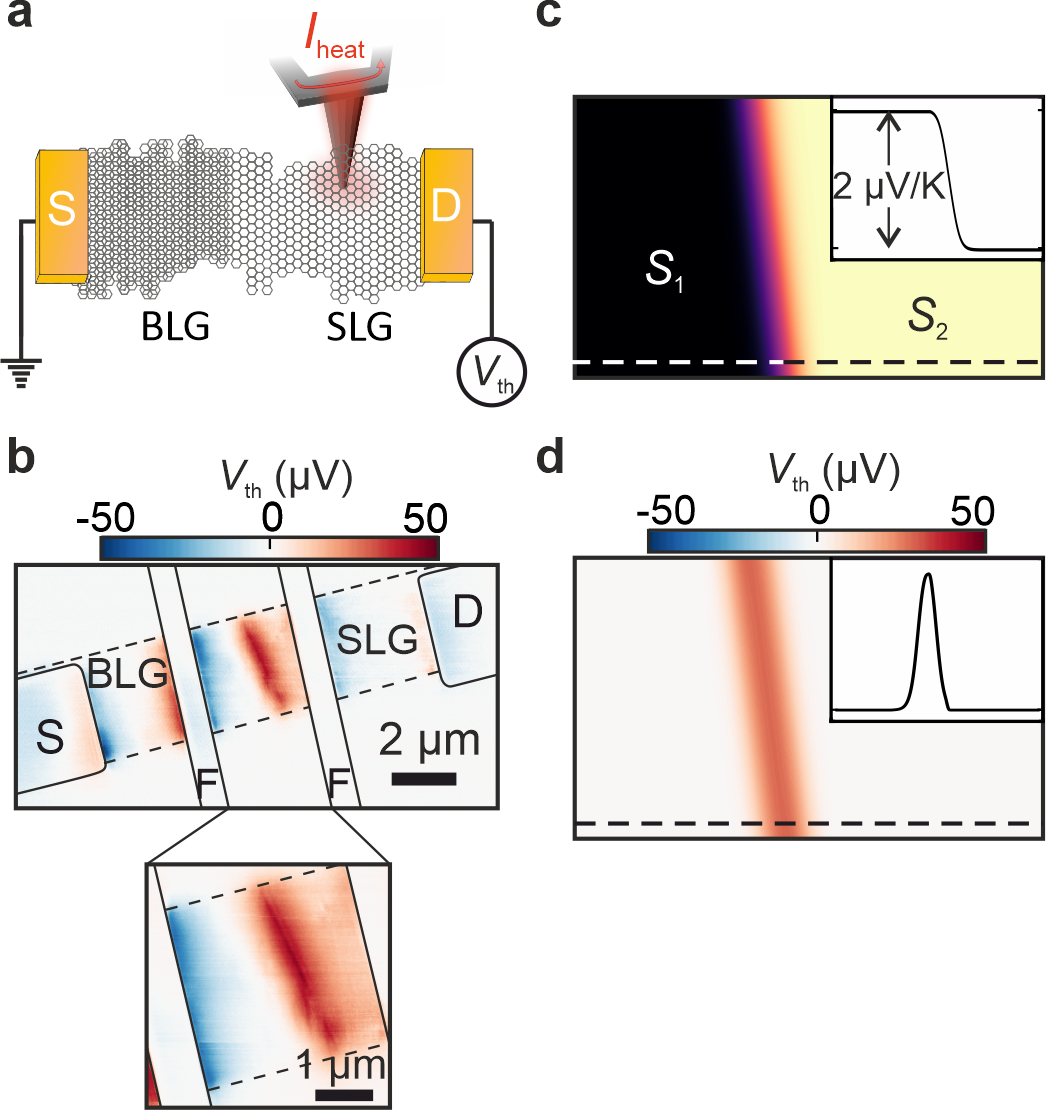}
	\caption{(a) Schematic of the STGM measurement setup. (b) Examplatory thermovoltage map of a SLG/BLG junction. The thermovoltage is measured between the source (S) and drain (D) contact with the other two contacts left floating (F). The dashed black lines indicate the position of the graphene flake and the black drawn through line the gold contacts. (c) Simulated Seebeck signal going from $S_\mathrm{1}$ to $S_\mathrm{2}$, resulting in $\Delta S = 2 \, \upmu$V/K. The inset shows a line trace through the Seebeck map with the position of the cut indicated by the dashed line. (d) Calculated thermovoltage map. The thermovoltage map was obtained by convoluting the Seebeck map in c) with the temperature distribution $dT(x)$ (see SI). The inset shows a line trace through the thermovoltage map with the position of the cut indicated by the dashed line. }
	\label{Fig1: SchematicDeconv}
\end{figure}
Here, we study graphene devices using Scanning Thermal Gate Microscopy (STGM), a novel non-invasive method to readily probe the local thermoelectric properties of devices made of 2D and thin film materials with few tens of nm lateral resolution. We first image a rectangular graphene strip as a function of gate voltage and secondly a more complex single-layer/bilayer graphene (SLG/BLG) junction device. Using STGM, we are able to resolve the impact of metallic contacts, changes in layer thickness and internal strain in the graphene and, importantly, can quantify the length scales involved with changes in the Seebeck coefficient due to different causes. While the developed method is applied to graphene devices here, it is generally suitable for any 2D or thin film device structure. \\

\section{Results and discussion}
\noindent Our devices are made from exfoliated single-layer and bilayer graphene top contacted by Ti-Au contacts on a Si/SiO$_2$ chip that serves as a global back gate (see methods for fabrication details). The measurements are performed employing a Scanning Thermal Microscope (SThM) setup, effectively an atomic force microscope (AFM) with a microfabricated resistor incorporated on the cantilever of the AFM probe close to the tip \cite{Gomes2015}. When applying a high DC or AC voltage to the resistor the tip can be heated up by tens of Kelvin above the ambient SThM temperature and used as a local heat source. As this heat source is scanned over the sample with nanometer precision, the position dependent open-circuit voltage drop on the device is recorded (see Figure \ref{Fig1: SchematicDeconv}a for the measurement schematic) and in equilibrium no current is flowing in the sample.
STGM effectively presents a three terminal probing technique where heat from the STGM tip modifies the potential voltage drop across the two electrical terminals, producing a 2D nanoscale map of the thermoelectric response. Crucially, STGM does not require an electrical contact between the tip and the probed sample to measure the local thermoelectric response.\\
For a given Seebeck coefficient $S(x)$ and a temperature increase caused by the tip $T(x)$, the tip position dependent thermovoltage can be written as (see SI)
\begin{equation}
V_\mathrm{th}(x_{\mbox{\tiny{T}}}) = -\int_{x_{\mbox{\tiny{L}}}}^{x_{\mbox{\tiny{R}}}} S(x) \frac{\partial T (x-x_{\mbox{\tiny{T}}})}{\partial x}dx~.
\label{Eq1:Vth}
\end{equation}
Here $x_{\mbox{\tiny{T}}}$ denotes the position of the tip, $x_{\mbox{\tiny{L}}}$ and $x_{\mbox{\tiny{R}}}$ the position of the left and right contact and $\frac{\partial T (x-x_{\mbox{\tiny{T}}})}{\partial x}$ the position dependent derivative of the temperature profile caused by the hot tip. Assuming that $T(x)$ is a symmetric function, Equation (\ref{Eq1:Vth}) shows that $V_\mathrm{th}$ reflects the local variations (asymmetry) of $S(x)$ around $x_\mathrm{T}$ within a length scale given by the thermal resistances of the sample. For a SLG/BLG junction for example, where symmetry is broken by the SLG/BLG step, we observe a positive thermovoltage signal at the junction when measuring the voltage drop at the SLG contact with the BLG contact grounded as shown in Figure \ref{Fig1: SchematicDeconv}b. The adjacent additional signal caused by the contacts will be discussed later. This positive thermovoltage signal can be explained by two different Seebeck coefficients in the single-layer and bilayer graphene area of the device, respectively, due to the different energy dispersion at low energies as previously reported \cite{Xu2010}. Assuming
a device where two materials with different Seebeck coefficients $S_1$ and $S_2$ are connected we can calculate the expected thermovoltage signal by convoluting $S(x)$ with $dT/dx$ as shown in Equation (\ref{Eq1:Vth}). The difference in Seebeck coefficient $\Delta S = S_2-S_1 = 2 \, \upmu$V/K is displayed in Figure \ref{Fig1: SchematicDeconv}c. Here the temperature distribution is assumed to have a Gaussian shape see (\cite{Hache2012} and SI) and we convolute each line of the Seebeck map with the temperature distribution to obtain the thermovoltage map. 
\begin{figure*}[ht]
	\centering
	\includegraphics[width=\linewidth]{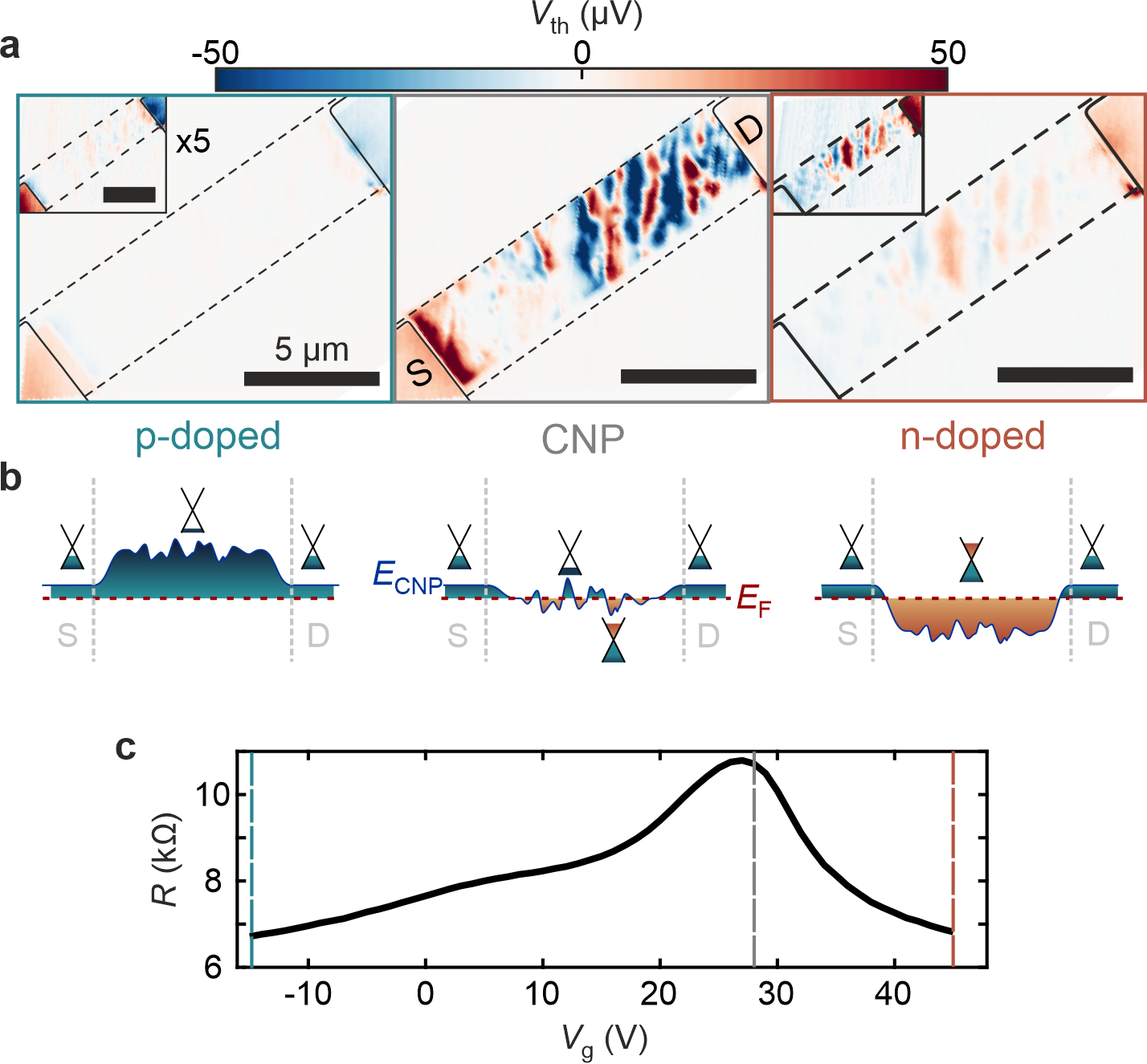}
	\caption{(a) Thermovoltage maps at three different gate voltages. The insets show the same map at a sensitivity of $V_\mathrm{th}$ -10 to 10 $\upmu$V and all scale bars denote 5 $\upmu$m. The dashed black lines indicate the position of the graphene flake and the black drawn through line the gold contacts. (b) Schematic showing the energy distribution in graphene relativ to the Fermi energy  band bending at the gold covered graphene to the graphene channel interface for a p-doped (left) and n-doped (right) graphene channel and at the charge neutrality point or CNP (middle). (c) Gatetrace of the device, showing the CNP point around 28 V. The dashed lines indicate the three gate voltages at which the above maps were recorded. }
	\label{Fig2: RectDevice}
\end{figure*}
As can be seen in Figure \ref{Fig1: SchematicDeconv}d, the resulting thermovoltage signal is positive along the position of the junction where the BLG changes to SLG, reproducing the experimentally observed results well. We note that the width of the thermovoltage peak in the thermovoltage map is given by the length scale on which the change in Seebeck coefficient from $S_1$ to $S_2$ happens, and the width of the temperature distribution (see SI). As discussed later, it is possible to extract the width of the temperature distribution from the thermovoltage maps, which yields information about the band bending when changing the graphene thickness.

\subsection{Rectangular Graphene Device}

\noindent We first measure the gate dependent local thermoelectric signal of a graphene rectangle (see Figure \ref{Fig2: RectDevice}). The charge neutrality point of the sample is at a back gate voltage of around $V_g = 28$ V, as inferred from the gate dependent measurement of the two-terminal channel resistance (see Figure \ref{Fig2: RectDevice}c). Thus, the graphene is strongly p-doped at $V_\mathrm{g} = 0$ V, which is due to the interaction of the SiO$_2$ substrate surface charges with the graphene as reported previously \cite{Shi2009,Wang2012}. 
An STGM thermovoltage map with a tip apex-temperature of $T_\mathrm{tip} \approx 53$ K is recorded at three back gate voltages: for a p-doped graphene channel at -15 V, close to the Dirac point at 28 V and for an n-doped graphene channel at 45 V (Figure \ref{Fig2: RectDevice}a, left to right). The tip-apex temperature is defined with respect to the microscope stage and when the tip is in contact with the sample surface.
In the p-doped regime, a thermovoltage on the order of 30 $\upmu$V is generated across the device when the hot tip heats up the areas around the electrical contacts, where the polarity of the induced thermovoltage is negative (positve) when the source (drain) is locally heated up. As before, the source is defined as the contact kept on ground during the measurement. Only very little to no thermovoltage drop is observed when the tip is scanned over the graphene channel. At low carrier concentrations around the Dirac point, a seemingly chaotic distribution of thermovoltage signal of changing polarity appears in the graphene channel. In the n-doped regime, the signal along the graphene strip is again reduced. A larger voltage build up (compared to that in	the channel) is observed when the source and drain contacts are heated up, with opposite polarity compared to the p-doped case.\\ 
The chaotic signal observed close to the Dirac point is attributed to charge puddles inside the graphene channel, which locally vary the energy dependent position of the charge neutrality point and are commonly observed for graphene on SiO$_2$ substrates (see Figure \ref{Fig2: RectDevice}b) \cite{Woessner2016,Xue2011}. This leads to sharp local variations in $S$ which in turn influences the thermovoltage signal. We note that the signal from the charge puddles dominates the map at the charge neutrality point. This is due to the high sensitivity of the magnitude and sign of the Seebeck coefficient on the carrier density around the Dirac point \cite{Woessner2016,Zuev2009}, resulting in large differences in $S$ over the short distances associated with charge puddles. \\
It is important to note, that a change in the thermal spreading resistance of the sample and the tip-sample interface thermal resistance, which is dominated by the thermal conductivity of the measured surface, strongly influences the tip-apex temperature. This is particularly relevant when the hot tip is in contact with the Au electrodes since the thermal conductivity of Au is two orders of magnitude larger than that of SiO$_2$/graphene on SiO$_2$. This leads to a reduction of the excess tip-apex temperature from $T_\mathrm{tip} \approx 53$ K to $T_\mathrm{tip} \approx 1.5$ K when the tip is in contact with Au compared to SiO$_2$/graphene \cite{Tovee2012}. 

\noindent In the following we discuss the origin of the observed thermovoltage signals. When the hot tip heats up the left or right electrical contact, the measured thermovoltage drop corresponds to the sum of the thermoelectric voltage build-up of the graphene under the Au contacts and the graphene channel. This signal will later be used to estimate the (global) Seebeck coefficient of the graphene channel with respect to the graphene underneath the contact \cite{Xia2009}. \\
The origin of thermoelectric signals is the preference of charge carriers to diffuse from hot regions in a material towards colder regions. The driving force for this process is proportional to the temperature difference $\Delta T$ and the entropy the system gains per diffusing carrier. The latter is highest if a carrier diffuses towards regions of higher DOS. This, depending on the source/drain configuration and the majority carriers in the sample, results in the build up of a positive or negative thermovoltage under open circuit conditions. \\
Graphene underneath the gold contacts is shielded from the electric field created by a gate voltage, such that its carrier concentration is not affected when a $V_\mathrm{g}$ is applied. 
For a gate voltage $V_\mathrm{g} = 45$ V, where the graphene channel is highly n-doped (see Figure \ref{Fig2: RectDevice}a right) the DOS at the Fermi energy is larger in the channel than in the graphene underneath the gold contact (see Figure \ref{Fig2: RectDevice}b right), meaning electrons will diffuse from the contact area into the graphene channel. This results in a positive (negative) thermovoltage build up at the drain (source) contact. While the DOS in the channel is still larger in the channel than in the graphene underneath the gold contact for a p-doped sample (see Figure \ref{Fig2: RectDevice}b left), the majority carriers are now holes, resulting in a negative (positive) thermovoltage build up at the drain (source) contact. 

\begin{figure*}
	\centering
	\includegraphics[width=\linewidth]{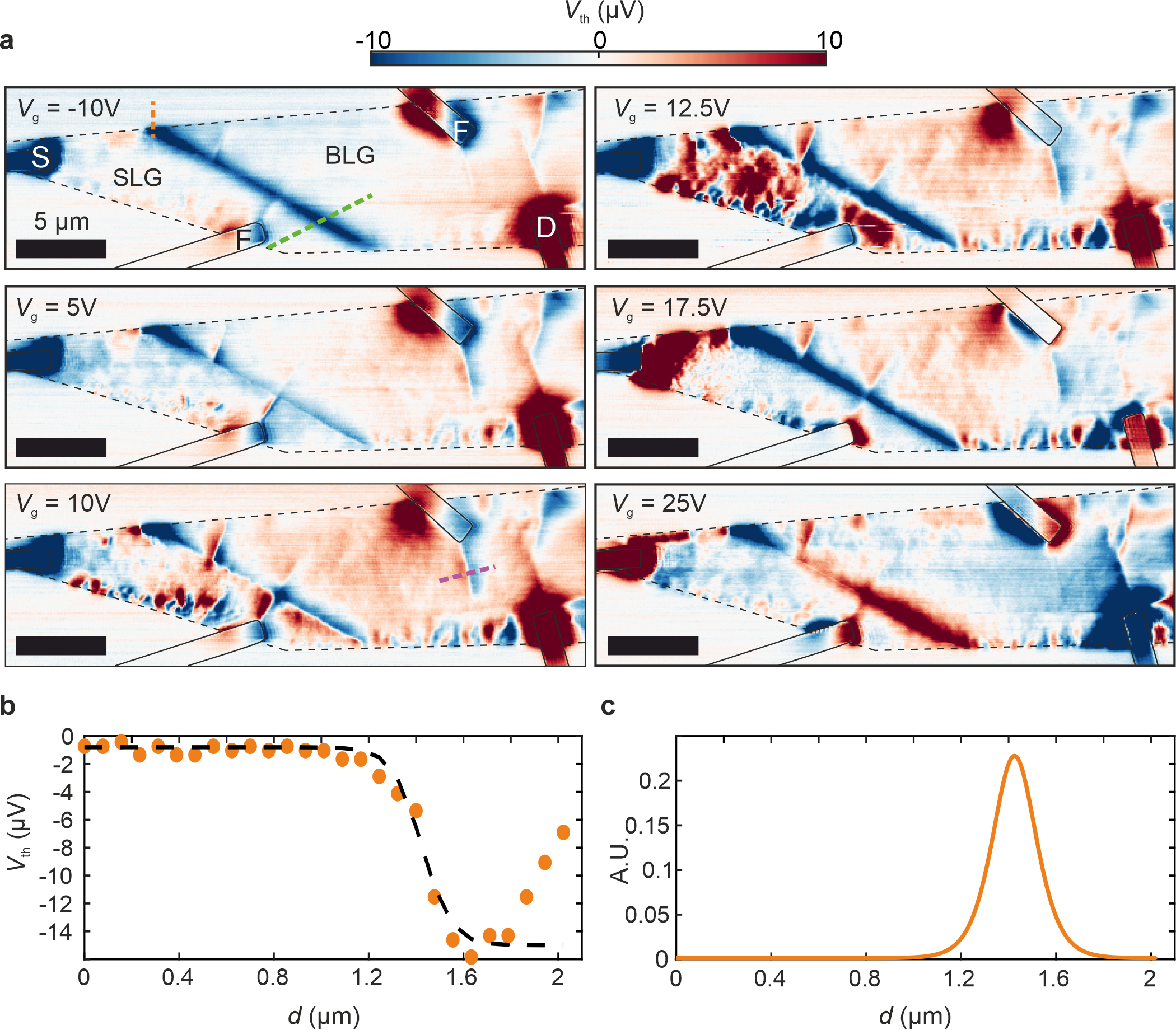}
	\caption{(a) Thermovoltage maps at six different gate voltages for a SLG and BLG graphene flake. The source and drain contacts are indicated by S and D respectively and the position of the line traces shown in Figure \ref{Fig4: DeconvTraces}a is indicated by the green dashed line in the top left and the line trace in Figure \ref{Fig4: DeconvTraces}c by the pink dashed line in the bottom left image. The dashed black lines indicate the position of the graphene flake and the black drawn through line the gold contacts. All scale bars denote 5 $\upmu$m. (b) Linecut indicated in the top left panel in a) by the orange dotted line (orange dots) and a sigmoidal Boltzman curve fit (black dotted line). (c) Derivative of the fit in b).  }
	\label{Fig3: Single-Bilayer}
\end{figure*}

\subsection{Single-layer/Bilayer Junction}
\noindent In the following, we apply STGM to a second type of graphene device, a junction formed between SLG and BLG (similar to the device and measurement shown in Figure \ref{Fig1: SchematicDeconv}b). Shown in Figure \ref{Fig3: Single-Bilayer}, the sample consists of a graphene flake that has both a single-layer and a bilayer section and is contacted by four gold contacts. The two contacts labelled source (S) and drain (D) are used for measuring the thermovoltage drop, while the other contacts are left floating (F). \\
In Figure \ref{Fig3: Single-Bilayer}, in accordance with the previous measurement of the graphene rectangle, a thermovoltage signal on the contacts is observed, which switches polarity as the gate voltage is swept over the charge neutrality point. The gate dependent thermovoltage maps highlight the impact of the contacts on the graphene sample. \\
The two passive, floating contacts show bi-polar signals, reversing polarity as the gate voltage is increased. This is because the floating contacts locally dope the graphene and pin the carrier density, an effect that has previously been observed in scanning photocurrent microscopy measurements \cite{Lee2008}. In addition, the Fermi level pinning from the contacts leads to a strongly non-homogenous change of the carrier density within the sample. Rather than occurring simultaneously over the whole sample area, the change is gradual, starting from the SLG/BLG junction and wandering out towards the contacts as can be seen for higher gate voltages (right side of Figure \ref{Fig3: Single-Bilayer} and further images in the SI). 
Furthermore, STGM allows us to image the gate screening due to a second layer of graphene on top of the SLG channel. While a multitude of charge puddles appear around the charge neutrality point for gate voltages $V_\mathrm{g} = 10/12.5$ V on the SLG side, the BLG area of the sample shows hardly any. This is attributed to the electric field created by surface dipoles in the oxide being screened by the bottom layer and consequently not affecting charge carriers in the top layer, as previously indirectly observed \cite{Sun2010,Ohta2007}. In addition, this observation provides evidence that the two graphene layers are thermally largely decoupled \cite{Rojo2019}. Assuming charge puddles are present in the bottom layer as they are in the SLG area, any temperature differential caused by the tip should cause a thermovoltage signal, which is not observed in our experiment. \\
Furthermore, we observe thermovoltage signals along straight lines (one example is highlighted by a pink dashed line in the bottom left panel of Figure \ref{Fig3: Single-Bilayer}a) which are not visible in the height topography AFM maps (see SI). This signal could be attributed to local strain in the graphene layer or wrinkles, induced by the deposition of the contacts or the annealing step. Strain has been shown to alter the band structure \cite{Yoon2011} and Seebeck coefficient in graphene as well as displaying a Peltier effect \cite{Lu2020}. Notably, at all gate voltages a strong thermovoltage signal is observed when heating up the interface between SLG and BLG as reported in scanning photocurrent measurements \cite{Mueller2009,Xu2010}. The signal is changing from negative to positive as the gate voltage is increased. Both the wrinkle and the SLG/BLG junction signal follow a non-linear gate dependence.\\ 
The change in the sign at the SLG/BLG interface with changing majority carrier concentration (here both the SLG and BLG area of the sample is assumed to be p-doped (n-doped) for low (high) gate voltages respectively) implies that there is a change in the Seebeck coefficient difference $\Delta S = S_\mathrm{SLG}-S_\mathrm{BLG}$. 

\begin{figure*}[ht]
	\centering
	\includegraphics[width=\linewidth]{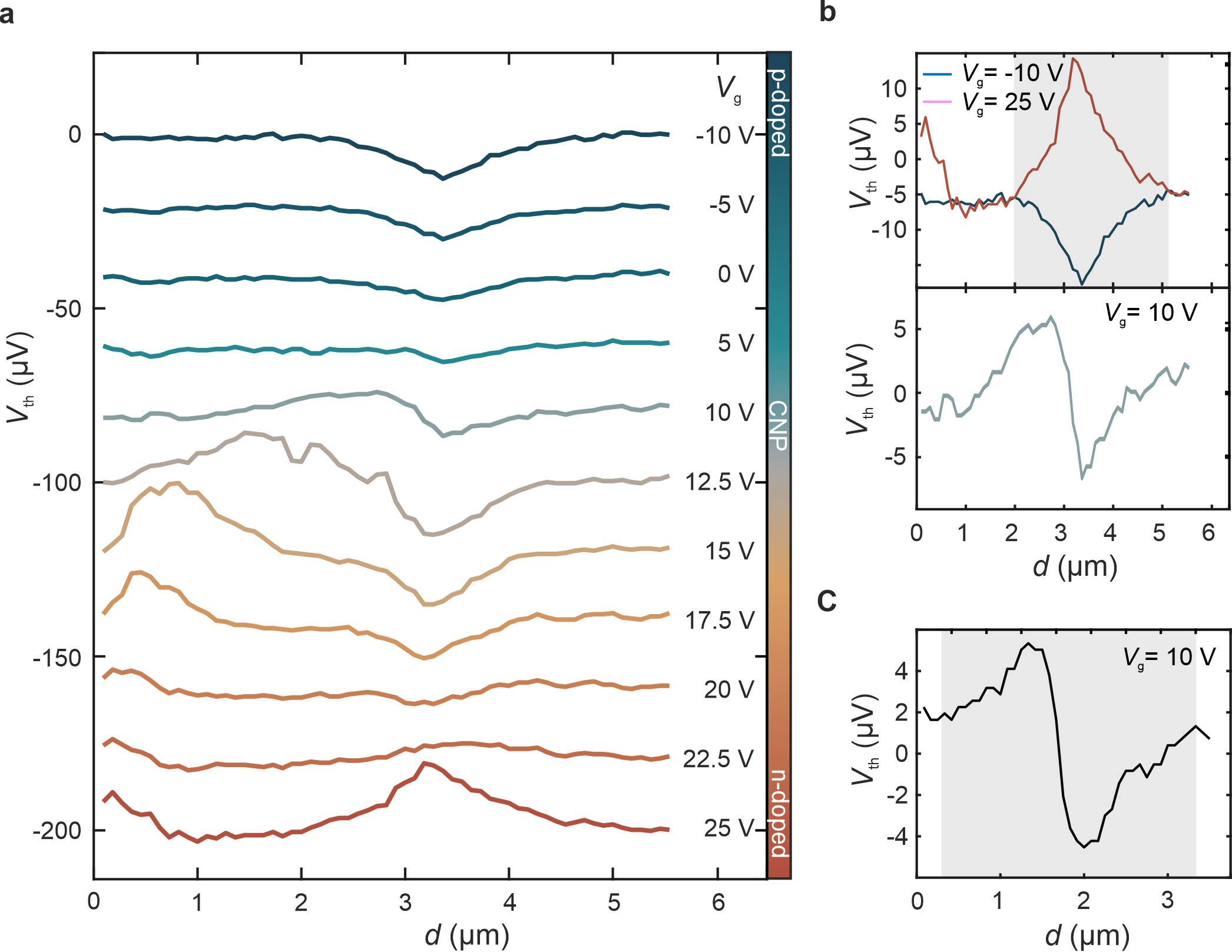}
	\caption{(a) Linetraces through the SLG/BLG junction shown in Figure \ref{Fig3: Single-Bilayer}a for different gate voltages. The linetraces are offset by -20 $\upmu$V each for clarity. (b) Linetrace through the SLG/BLG junction at $V_\mathrm{g} = -10$ V and $V_\mathrm{g} = 25$ V showing a dip/peak (top) and at $V_\mathrm{g} = 10$ V showing a bipolar signal (bottom). (c) Linecut through the thermovoltage signal of a wrinkle in the SLG/BLG sample. The grey shaded area indicates the width of the signal.}
	\label{Fig4: DeconvTraces}
\end{figure*}

\noindent As discussed above, diffusion of carriers is driven by a gain in entropy, and carriers diffuse from regions of low DOS into regions of high DOS. Due to the parabolic dispersion relation in BLG compared to the linear one in SLG around the charge neutrality point, the DOS in BLG is higher than in SLG \cite{McCann2013,CastroNeto2009}. This leads to a positive (negative) thermovoltage build-up for electron (hole) doped graphene when measuring the voltage at the drain contact with respect to the source \cite{Xu2010}. This development is shown in Figure \ref{Fig4: DeconvTraces}a where line traces along the green dotted line in Figure \ref{Fig3: Single-Bilayer}a through the junction at different gate voltages are displayed. Figure \ref{Fig4: DeconvTraces}b, top, shows close ups of the position dependent thermovoltage signal for a high p-doped ($V_\mathrm{g} = -10$ V) and n-doped ($V_\mathrm{g} = 25$ V) sample. For p-doped (n-doped) graphene, the signal dips (peaks) at the interface between SLG and BLG. Such a dip (peak) observed for the highly p-doped (n-doped) sample, respectively, is predicted by the model presented in Figure \ref{Fig1: SchematicDeconv} and can be attributed to a step-wise change in the Seebeck coefficient between the SLG and BLG areas. 

\noindent However, at gate voltages close to the charge neutrality point of the sample ($V_\mathrm{g} = 10/12.5$ V) the thermovoltage exhibits a bipolar signal not compatible with a Seebeck coefficient changing from $S_\mathrm{1}$ to $S_\mathrm{2}$. A similar bipolar signal at $V_\mathrm{g} = 10$ V is found around wrinkles as shown for an exemplary linecut in Figure \ref{Fig4: DeconvTraces}c. Rather than indicating a transition from one Seebeck coefficient value to another, a bipolar signal implies a drop/peak of the Seebeck coefficient (see SI for simulations similar to the ones shown in Figure \ref{Fig1: SchematicDeconv}). For the wrinkles, the sudden spatial change in the Seebeck coefficient can be explained by local strain, which is the origin of the fold or wrinkle formation in the first place. It has been predicted that local strain in graphene can alter its Seebeck coefficient with the magnitude of this effect being largest around the charge neutrality point \cite{Nguyen2015}. As a result, wrinkles in graphene exhibit a bipolar and prominent signal around the CNP. At higher doping, the change in the Seebeck coefficient is not as large and only leads to a small $\Delta S$. While this describes the signal at the wrinkles, the nature of the bipolar thermovoltage displayed at the SLG/BLG junction is less clear. The bipolar thermovoltage could be a combination of the signal from charge puddles and the junction or be attributed to local symmetry breaking and disorder at the edges of the BLG which locally influences the Seebeck coefficient. \\
As explained above, the magnitude of the local thermovoltage signal measured by STGM is given by a convolution of a temperature difference with a given spatial Seebeck coefficient distribution. Thus it is possible to extract the length scales the Seebeck coefficient is changing on associated with the different origins of the signals. We note that the spatial distribution in the thermovoltage signals is directly influenced by two factors: the shape and height of the temperature profile distribution created by the STGM tip and the local changes of the Seebeck coefficient. In order to determine the width of the temperature distribution created by the STGM tip, we fit the step in the thermovoltage signal at the graphene flake edge with a sigmoidal Boltzmann curve $f(x) = y_1+(y_1-y_2)/(1_e^{\frac{x-x_0}{\Delta x}})$ (see orange dashed line in the top left pannel in Figure \ref{Fig3: Single-Bilayer}a and SI) \cite{Tovee2014}. This fit gives $\Delta x \approx 60$nm, suggesting a lateral resolution of $2 \Delta x$ = 120nm. From this fit we can also extract a width of the Gaussian temperature distribution of $l_\mathrm{dT} \approx 1$ $\upmu$m (see SI). \\
By comparing $l_\mathrm{dT}$ to the length scales observed in the experiment for the thermovoltage signal, $l_\mathrm{V} \geq 3 \, \upmu$m (see Figure \ref{Fig4: DeconvTraces}b and c), it becomes clear that the width of the signal can not be explained by limited spatial resolution. Since the thermovoltage signal measured is a convolution of $S$ and $dT$, the signal width is approximately given by the sum of $l_\mathrm{dT}$ and $l_\mathrm{S}$ (the length scale associated with changes in the Seebeck coefficient), making it possible to extract $l_\mathrm{S}$. The signal width can then be used to quantify the band bending at the transition between SLG to BLG (similarly to the depletion region width in a semiconductor p-n junction that is determined by doping levels and applied voltage). 
Applying this to the line traces shown in Figure \ref{Fig4: DeconvTraces}a which show $l_\mathrm{V} \approx 3$ $\upmu$m, we can extract the length scale of the Seebeck coefficient transition when moving from SLG to BLG, giving $l_\mathrm{S} \approx$ 2 $\upmu$m for both a p-doped and an n-doped sample. We note that our results suggest that $l_\mathrm{S}$ depends on the carrier density: by lowering the carrier density we find that $l_\mathrm{S}$ decreases to $l_\mathrm{S} \approx$ 1 $\upmu$m at $V_\mathrm{g} = 5$ V (see Figure \ref{Fig4: DeconvTraces}a and SI). It is worth to mention that, although the length scales on which the Seebeck coefficient changes around the SLG/BLG junction and a wrinkle are similar, the underlying mechanism for this length scale is different (see Figure \ref{Fig4: DeconvTraces}b top and c). While $l_\mathrm{S}$ around the junction depends on charge carrier density,  $l_\mathrm{S}$ for wrinkles in graphene is expected to depend more on the elastic properties of graphene. Thus the above analysis suggests that even for an atomically sharp step between two 2D materials, the Seebeck coefficient changes gradually on a length scale of micrometers.

\subsection{Estimating the Seebeck coefficient}

\noindent Following Equation (\ref{Eq1:Vth}), the local thermovoltage is a convolution of the local variation of the Seebeck coefficient and the temperature gradient induced by the STGM tip. Therefore it is possible to extract the local Seebeck coefficient variations from a measurement of the local thermovoltage by deconvolution. We test this idea by making the following assumptions: 1) we assume a Gaussian distribution of the temperature profile induced by the heated tip, with $\sigma \approx 100$nm as extracted from the sigmoidal Boltzmann curve fit (see SI). 2) we assume a constant thermal spreading resistance and therefore a constant tip-apex temperature. The error making this assumption is low if only the graphene channel is investigated and contact areas are neglected. 3) the thermovoltage is measured between two sample contacts in the x-direction and therefore contributions in the y-direction can be neglected (here we rotate the thermovoltage maps before performing the deconvolution). \\ 

From Equation (\ref{Eq1:Vth}) we know that $V_\mathrm{th}$ is the convolution of $S$ and the position dependent derivative of the temperature profile caused by the hot tip $T'(x_{\mbox{\tiny{T}}})$, i.e. $V_\mathrm{th}(x_{\mbox{\tiny{T}}}) = (S*T')(x_{\mbox{\tiny{T}}})$. Using the temperature distribution $T(x_{\mbox{\tiny{T}}})$, and thereby $T'(x_{\mbox{\tiny{T}}})$, we can then deconvolute Equation (\ref{Eq1:Vth}) and obtain an estimate of the Seebeck coefficient distribution along the sample, $S(x)$ (see SI). Using a regularized filter in combination with a constraint least-square restoration algorithm to perform the 1D deconvolution, a quasi two-dimensional temperature distribution is line-by-line deconvoluted with the thermovoltage map to produce the Seebeck map (see methods and SI). \\ 
As the tip is moved over the sample, the centre of its two-dimensional temperature distribution moves as well. Then, at every point of the thermovoltage map, a complete deconvolution of each line in the temperature distribution with the thermovoltage image gives a point in the deconvolution image. 

\begin{figure}
	\centering
	\includegraphics[width=\linewidth]{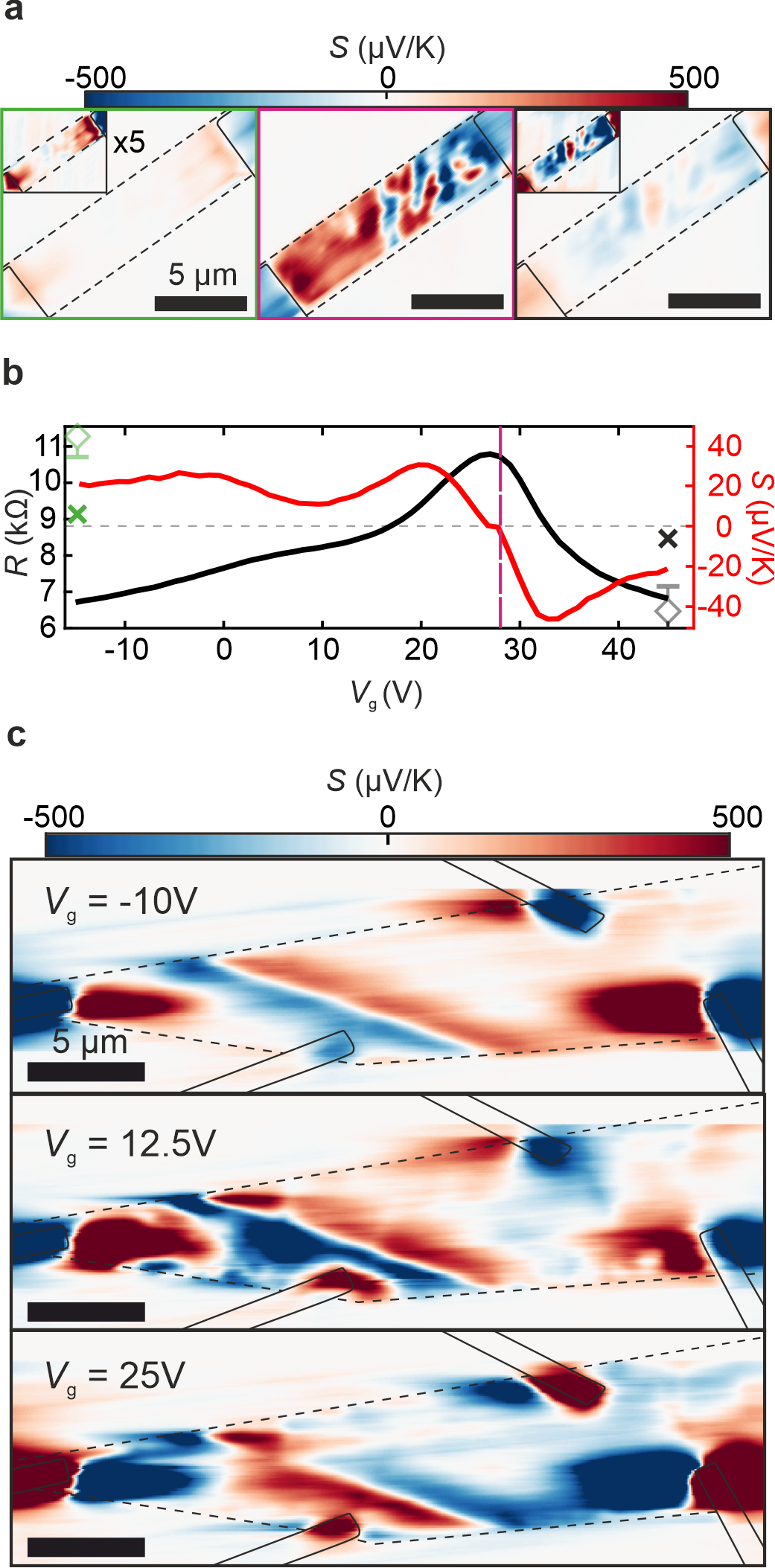}
	\caption{(a) Deconvolution of the thermovoltage maps shown in Figure \ref{Fig2: RectDevice}a. The insets show the same map at a sensitivity of $S$ -20 to 20 $\upmu$V/K. (b) Gatetrace of the rectangular device (black line), Seebeck coefficient calculated from Equation (\ref{Eq2:Mott}) (red line), Seebeck coefficient estimated from the thermovoltage maps (semitransparent diamonds) and the Seebeck coefficient extracted out of the deconvolution in a) (crosses). (c) Deconvolution of the thermovoltage maps shown in Figure \ref{Fig3: Single-Bilayer}. All scale bars denote 5 $\upmu$m. The dashed black lines show the position of the graphene flake and the black drawn through line the gold contacts.}
	\label{Fig5: DeconvS}
\end{figure}

\noindent The resulting deconvolution is shown in Figure \ref{Fig5: DeconvS}a. First, we note that the signal on and around the contacts is not accurate as we assumed a constant excess temperature of the tip-sample contact (of $T_\mathrm{tip}\approx$ 53 K) for the deconvolution. This is not given on the Au contacts where the tip-sample contact temperature drops substantially as discussed earlier. Therefore, the areas of the Au contacts, which effectively show the difference in Seebeck coefficients between gold and graphene \cite{Vera-Marun2016}, do not necessarily have the right magnitude (see discussion below). The graphene channel in between the contacts however, satisfies assumption 2 and displays the expected change in the Seebeck coefficient from positive to negative with increasing gate voltage \cite{Zuev2009}. In addition, large local variations of the magnitude and sign of the local Seebeck coefficient are observed around the Dirac point (see Figure \ref{Fig5: DeconvS}). These give rise to the thermovoltage signal shown in Figure \ref{Fig2: RectDevice}a.\\
\noindent In order to evaluate the deconvolution method presented in this work, we can compare its result to the Seebeck coefficient of the graphene channel obtained via two additional methods: 1) using the gate dependent conductance data of the device and the Mott formula to estimate $S$ \cite{Zuev2009} and 2) estimating the temperature drop $\Delta T$ over the whole sample when the STGM tip heats up one electrical contact, and use $S = V_\mathrm{th}/\Delta T$. 
For case 1), the Mott formula is given by
\begin{equation}
S = -\frac{\pi^2k_B^2T}{3 |e|}\frac{1}{G}\frac{dG}{dV_g}\frac{dV_g}{dE}\big|_{E=E_F} ~,
\label{Eq2:Mott}
\end{equation}
where $e$ is the electron charge, $G$ the electrical conductance, $k_\mathrm{B}$ the Boltzman constant and $T$ the operating temperature (see SI). The gate dependent Seebeck coefficient of the rectangular sample can then be calculated using the gatetrace from Figure \ref{Fig2: RectDevice}c with the results of this calculation shown in Figure \ref{Fig5: DeconvS}b. The calculated Seebeck coefficient overall first increases, changes sign when crossing the charge neutrality point, further decreases, and subsequently increases towards zero again. We observe maximum and minimum values of the Seebeck coefficient around $S$ = 30-40 $\upmu$V/K and $S$ = -40 to -50 $\upmu$V/K, respectively.\\
Then, for method 2), we divide the average thermovoltage value on the contacts in Figure \ref{Fig2: RectDevice}a by the assumed temperature difference between the contacts ($\approx$ 1.5 K as discussed) giving $S$ = 5.92 $\pm 3.2$ $\upmu$V/K and $S$ = -4.19 $\pm 1.49$ $\upmu$V/K for the Seebeck coefficient at low and high gate voltage respectively. The distribution of values of $V_\mathrm{th}$ on the contacts is plotted in a bar graph and fitted with a multi peaked Gaussian to determine the most likely event with the error then given by the width of the Gaussian (see SI). \\
Lastly, the Seebeck coefficient obtained via the deconvolution is extracted from the graphene channel in the maps of the spatial distribution of the Seebeck coefficient in Figure \ref{Fig5: DeconvS}a. Again, the values of $S$ in the channel are found by fitting the distribution with a multi peaked Gaussian to determine the most likely event with the error then given by the width of the Gaussian (see SI). This gives $S$ = 44.94 $\pm 5.43$ $\upmu$V/K and $S$ = -42.91 $\pm 9.4$ $\upmu$V/K for the p-doped and n-doped channel respectively. The result of the three different methods to obtain the Seebeck coefficient in the sample are shown in Figure \ref{Fig5: DeconvS}b. 
The values found for all methods are qualitatively on the same order, considering the assumptions necessary. In addition, all methods find the same trend of a reversal in sign when changing majority carriers as expected for graphene \cite{Zuev2009}. Method 2), that is the Seebeck coefficient estimated via the thermovoltage drop, quantitatively differs most likely due to the large error in determining the temperature difference across the device from heating up one contact.\\
The deconvolution can also be applied to the more complex thermovoltage maps of the SLG/BLG sample (Figure \ref{Fig5: DeconvS}c). Here, artefacts introduced by the changing tip-sample thermal resistance are clearly visible in the contact area. However, trends predicted by theory like the change in the Seebeck coefficient of the SLG area $S_\mathrm{SLG}$ with respect to the BLG area $S_\mathrm{BLG}$ are visible, with $S_\mathrm{SLG} < S_\mathrm{BLG}$ at $V_\mathrm{g} = -10$ V and $S_\mathrm{SLG} > S_\mathrm{BLG}$ at $V_\mathrm{g} = 25$ V \cite{Xu2010}. In addition, again, a high local variation of $S_\mathrm{SLG}$ with changing polarity can be seen around the charge neutrality point, caused by charge puddles. While the calculated Seebeck coefficients around 500 $\upmu$V/K may seem high for a graphene channel, we note that these are local variations due to local effects rather than representations of the global Seebeck coefficient. 

\section{Conclusions}
\noindent To summarize, we used STGM, a novel method to probe the local thermoelectric properties of two-dimensional and thin-ﬁlm devices on a nanometre scale. 
STGM is a mixed-physics approach where the voltage response on the two electrical terminals of the device is mapped as a function of the position and the temperature of the nanoscale probed scanned across the surface of the studied nanostructure.
We were able to investigate the thermoelectric fingerprint of charge puddles in graphene, resolve effects in our sample such as Fermi level pinning by the metallic contacts leading to non-homogenous carrier concentration distribution and reveal how the local carrier concentration changes when a gate voltage is applied. Furthermore, we find that strain strongly impacts the local Seebeck coefficient in single-layer and multilayer graphene sheets. The STGM method developed in this paper further allows us to obtain the local Seebeck coefficient through deconvolution with a good qualitative agreement between values obtained by deconvolution and theoretical prediction.\\
STGM provides valuable insights in the thermoelectric phenomena in two-dimensional materials by noninvasively studying planar junctions and the influence of local defects or metal contacts on the thermoelectric properties of the sample. STGM provides important insights leading to new strategies to increase the efficiency of thermoelectric devices and limit thermoelectric noise in electronic devices. STGM is not restricted to 2D materials but could also be applied to study the local thermoelectric properties around single molecule junctions or quantum dot devices \cite{Harzheim2020, Gehring2019,Gehring2017}.

\section{Experimental}

\noindent \textbf{Device Fabrication.}
The devices were fabricated by exfoliating HOPG on top of a standard Si/SiO$_2$ chip with a 300 nm oxide layer. Subsequently, if required, the graphene was patterned employing standard electron-beam lithography and then etched using oxygen plasma. Then, 1 nm Ti/50 nm Au contacts were deposited on top of the graphene and the final devices were annealed at 350 degree Celsius under a forming gas atmosphere for 30 minutes. \\
\textbf{Scanning Thermal Gate Measurements.}
The SThM is located in a high vacuum environment, prohibiting parasitic heat transfer as well as the formation of a water meniscus between the tip and the sample to achieve a better thermal resolution \cite{Menges2016,Kim2012}. In our measurements, the spatial resolution is limited by the size of the tip–sample contact, which is on the order of tens of nanometers. \\
For the STGM measurements, the SThM tip is heated up by applying a high AC voltage of $V_{heater} = 4$ $V_\mathrm{pp}$ at a frequency of 91 kHz for the thermal conductance measurements with a 4 V DC offset to the temperature sensor. This causes Joule heating, which results in an SThM tip excess temperature of $T_\mathrm{tip} \approx 53$ K at the interface with graphene on the SiO$_2$/Si substrate (see Supporting Information). This local heat source is then scanned over the sample while the global voltage drop $V_\mathrm{th}$ over the two contacts is measured with a SRS650 low-noise preamplifier and a SR560 high-impedance voltage preamplifier. The thermovoltage measurements do not require electrical contact between the tip and the sample, and thereby eliminate linked uncertainty, as well as requirements on the strength of the electrical tip–sample contact \cite{Lee2012}. The SThM can also be used in a passive tip configuration with the temperature of the sample modified via cooling (heating) of the stage.\\ We calibrate the electrical power applied to the tip resistor as a function of temperature on a heated stage inside a high-vacuum chamber as described elsewhere \cite{Tovee2012,Evangeli2019}. Crucially, here the tip apex temperature is a fraction of the cantilever temperature depending on the thermal conductivity of the material in contact (See SI for more details). \\
Offsets introduced by the amplifier are accounted for by subtracting the background signal far away from the sample from the whole map.\\
\textbf{Deconvolution.} 
The Deconvolution is performed using a regularized filter algorithm rather than one based on the more straight forward polynomial division, as the latter can become numerically unstable for large denominator coefficients. The filter used is a constrained least-squares filter, which aims to minimize the difference between the ideal and the restored image while keeping the high frequency and thereby noise component to a minimum \cite{Hunt1973,Lagendijk2009}. The point-spread function used for deconvolution is the derivative of the Gaussian temperature distribution (Figure \ref{Fig3: Single-Bilayer}c and SI), and a Laplacian operator is used as the regularization operator to retain image smoothness.\\

\begin{acknowledgments}
	OVK achnowledges EU project QUANTIHEAT (Grant 604668), EPSRC project EP/K023373/1, UKRI project NEXGENNA and Paul Instrument Fund, c/o The Royal Society. P.G. acknowledges a Marie Skłodowska-Curie Individual Fellowship under Grant TherSpinMol (ID: 748642) from the European Union’s Horizon 2020 research and innovation programme. CE aknowledges a Marie Sklodowska-Curie Individual Fellowship under Grant NOSTA (ID: 704280) and EPSRC grant EP/N017188/1.
\end{acknowledgments}

\clearpage

\onecolumngrid

\setcounter{equation}{0}
\setcounter{figure}{0}
\renewcommand{\theequation}{S\arabic{equation}}
\renewcommand{\thefigure}{S\arabic{figure}}

\large{\textbf{SI: Direct Mapping of Local Seebeck Coefficient in 2D Material Nanostructures via Scanning Thermal Gate Microscopy}}

\section{Thermovoltage gate dependence}
\begin{figure}[h]
	\centering
	\includegraphics[width=\linewidth]{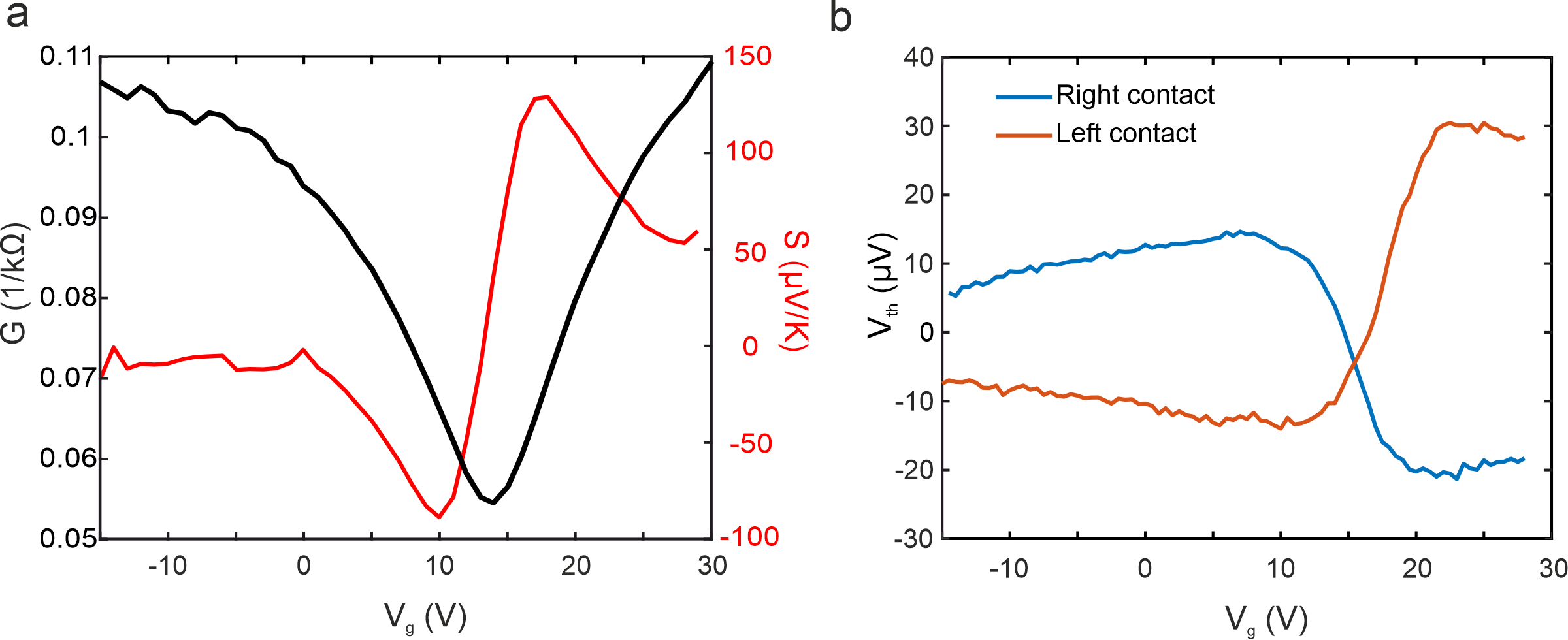}
	\caption{(a) Gate dependent conductance (black) and calculated (red) Seebeck coefficient using the Mott formula for an additional device. (b) Left (blue) and right (red) contact signal as a function of gate voltage. }
	\label{FigSI1: Gatedep}
\end{figure}

In order to illustrate the gate dependence of the thermovoltage, another SLG graphene device with higher gate resolution is studied. Figure \ref{FigSI1: Gatedep}a shows the conductance gate trace of the device and the calculated Seebeck coefficient using the Mott formula from Equation (2) in the manuscript. In \ref{FigSI1: Gatedep}b the gate dependence of the thermovoltage is shown, exhibiting a clear change from positive to negative for one contact and vice versa for the other one. This lineshape matches previous measurements very well and indicates a change in the majority carrier as reflected in the conductance trace. We note, that the signals change sign at slightly different gate voltages (though both in the vicinity of the Dirac point shown in \ref{FigSI1: Gatedep}a), most likely due to slight variations in the gating of the graphene underneath the contacts. 

\section{Temperature Tip calibration}

The temperature of the tip apex is a fraction of the tip-resistor temperature due to the conical shape of the tip, its material and the material the tip is in contact with. For a low thermal conductivity material with poor heat dissipation the temperature at the tip-sample contact will be higher than for a material with a higher thermal conductivity. The tip apex temperature when in contact with different materials can be found in \cite{Tovee2012,Evangeli2019} as obtained by experiments and COMSOL simulations.
For example, the temperature increase at the tip when in contact with gold (thermal conductivity of 310 W$\mathrm{m^{-1}K^{-1}}$) is $\approx$ 3-5\% of the heater temperature and for 300 nm of SiO$_2$ (thermal conductivity of about 1 W$\mathrm{m^{-1}K^{-1}}$) it is $\approx$ 80-85\% of the heater temperature.

\section{Pseudo 2D Deconvolution}

\begin{figure}[h]
	\centering
	\includegraphics[width=\linewidth]{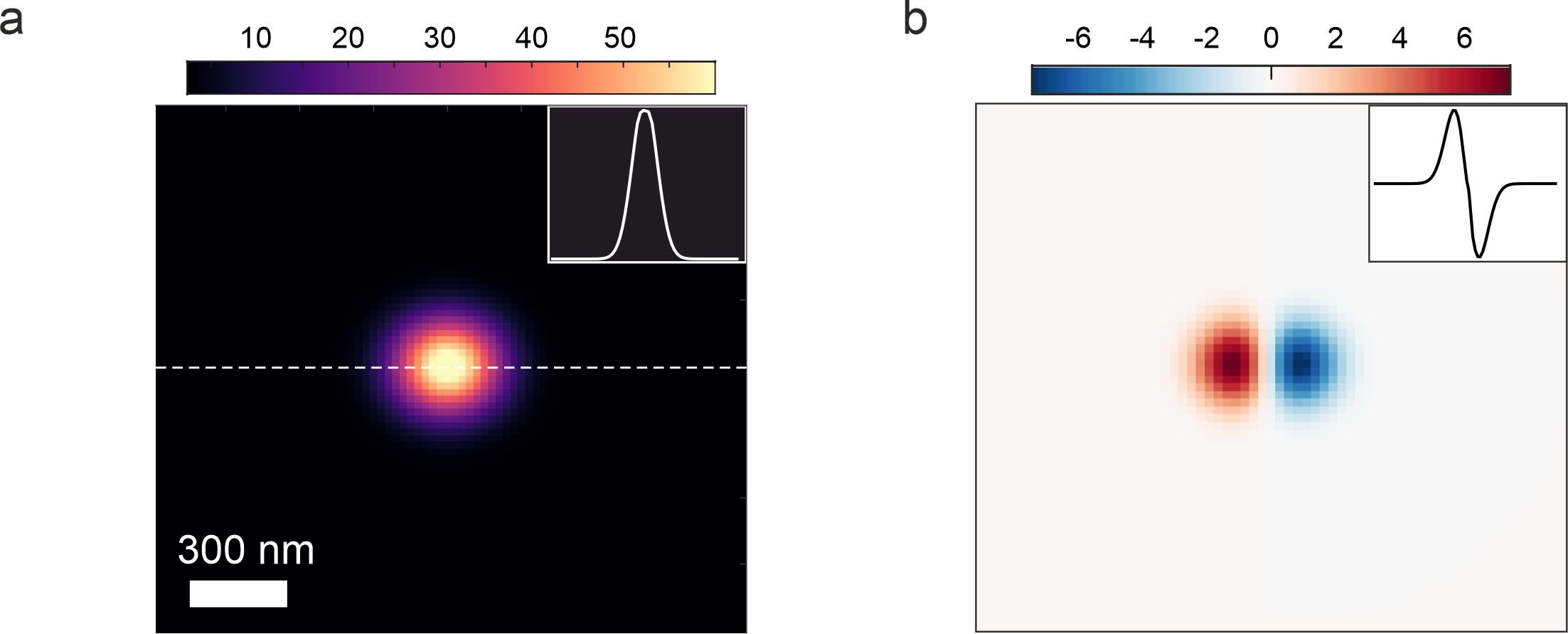}
	\caption{(a) Temperature distribution based on a Gaussian function, insets shows the linecut through the white dashed line. (b) Temperature distribution in a) differentiated with respect to the x-axis, insets shows the linecut through the black dashed line. }
	\label{FigSI2: TempDistr}
\end{figure}

In order to deconvolute the thermovoltage maps and extract the Seebeck coefficient, a pseudo 2D deconvolution is used. First, a 2D map of the assumed temperature distribution on the sample, a Gaussian function, is calulated (see Figure \ref{FigSI2: TempDistr}a).

\begin{figure}[h]
	\centering
	\includegraphics[width=\linewidth]{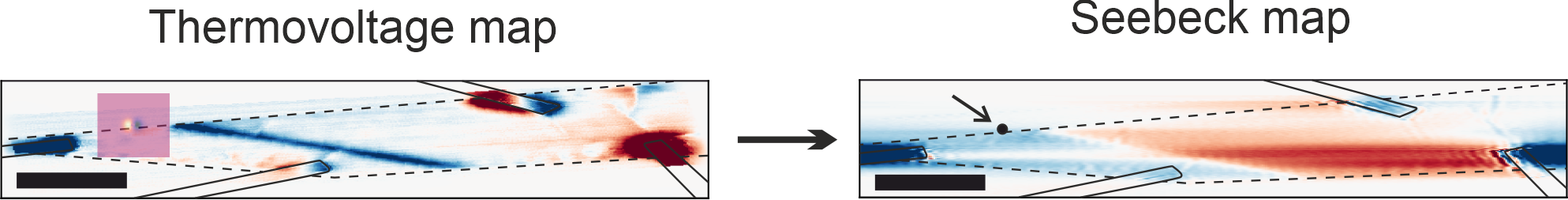}
	\caption{(a) Schematic showing the deconvolution of the temperature gradient (semi-transparent overlaid on the thermovoltage map) with the thermovoltage map (left) and the resulting pixel in the Seebeck map (marked by the black arrow in the Seebeck map).}
	\label{FigSI3: Pseudo2Ddeconv}
\end{figure}

In a second step, this temperature distribution is differentiated along the x-axis as shown in Figure \ref{FigSI2: TempDistr}b. The differentiation is along the x-axis since we align all thermovoltages such that the two measurement contacts are aligned along the x-axis. We then assume any distribution to the thermovoltage that is resulting from the temperature gradient in the y-direction does not contribute to the measured thermovoltage and is neglected in the following.\\
In order to calculate the 2D Seebeck coefficient, this temperature gradient is then transposed on top of part of the thermovoltage image, that is with each line corresponding to one line in the thermovoltage image (see Figure \ref{FigSI3: Pseudo2Ddeconv}). Then each line of the temperature gradient image is deconvoluted with each line of the thermovoltage image in a 1D line convolution described in the methods and the result added up. Since here the middle of the temperature gradient image represents the position of the heated SThM tip, this gradient image is then scanned over the thermovoltage image, with each pixel in the resulting Seebeck picture corresponding to the deconvolution of the surrounding area around the respective pixel in the thermovoltage with the temperature gradient image.  
The algorithm used here is deconvreg which is part of the Matlab image processing toolbox and uses a regularized filter algorithm. It assumes, that the image to be deconvoluted (the thermovoltage map) was created by convolving the original image (the Seebeck coefficient map) with a point spread function (the temperature profile of the SThM tip) and the addition of noise. Crucially, since this algorithm is developed for images, it does not take into account the dimension of the original image or the deconvolved image. We will examine below how to correct for this. \\

For two recorded images ($f$ and $g$) that are physical fields (that is moving along the pixels in the x and y direction of the image is associated with a change in a physical quantity such as position), their deconvolution $h$ is again a physical field. The units of the deconvolution are then given by $dim(h) = dim(f)dim(g)dim(A)$ where $dim(A)$ is the dimension of the image area, usually square meters. The convolution $h = f*g$ can then be written as a discrete Riemann sum as
\begin{equation}
h_n \approx \Delta A \sum_{k}f_{n-k} g_k ~,
\end{equation}
where the dimensionality of the image is taken into account by $\Delta A$, the area of one pixel. In digital image processing, the units of the image are often ignored (as is the case in our deconvolution algorithm), meaning the convolution in this case is just given by 
\begin{equation}
h_n \approx \sum_{k}f_{n-k}g_k ~.
\end{equation} 
As a result, in order to arrive at the right expression for the deconvolution we have to take into account the dimensionality of the original image by multiplying our image $h_n$ by $\Delta A$. The algorithm used here is essentially employing Fourier transforms to perform the deconvolution, that is the deconvolution is given by:
\begin{equation}
f = \mathrm{ifft}(F) = \mathrm{ifft}(H/G) ~,
\end{equation}
where $F,G$ and $H$ are the Fourier transforms of $f,g$ and $h$ respectively and $\mathrm{ifft()}$ denotes the inverse Fourier transform. Since Fourier transforms and therefore its inverse obey linearity the deconvolution, taking into account the dimensionality via $\Delta A$, is given by
\begin{equation}
f = \mathrm{ifft}(H/G) \Delta A ~.
\end{equation}
Therefore, we need to multiply the resulting deconvolved Seebeck map by $\Delta A$, which is equal to 0.0022 $\upmu m^2$ for the device shown in Figure 2 and 0.0047 $\upmu m^2$ for the device shown in Figure 3.

\section{Additional Thermovoltage measurements on the SLG/BLG device}

\begin{figure}[h!]
	\centering
	\includegraphics[width=\linewidth]{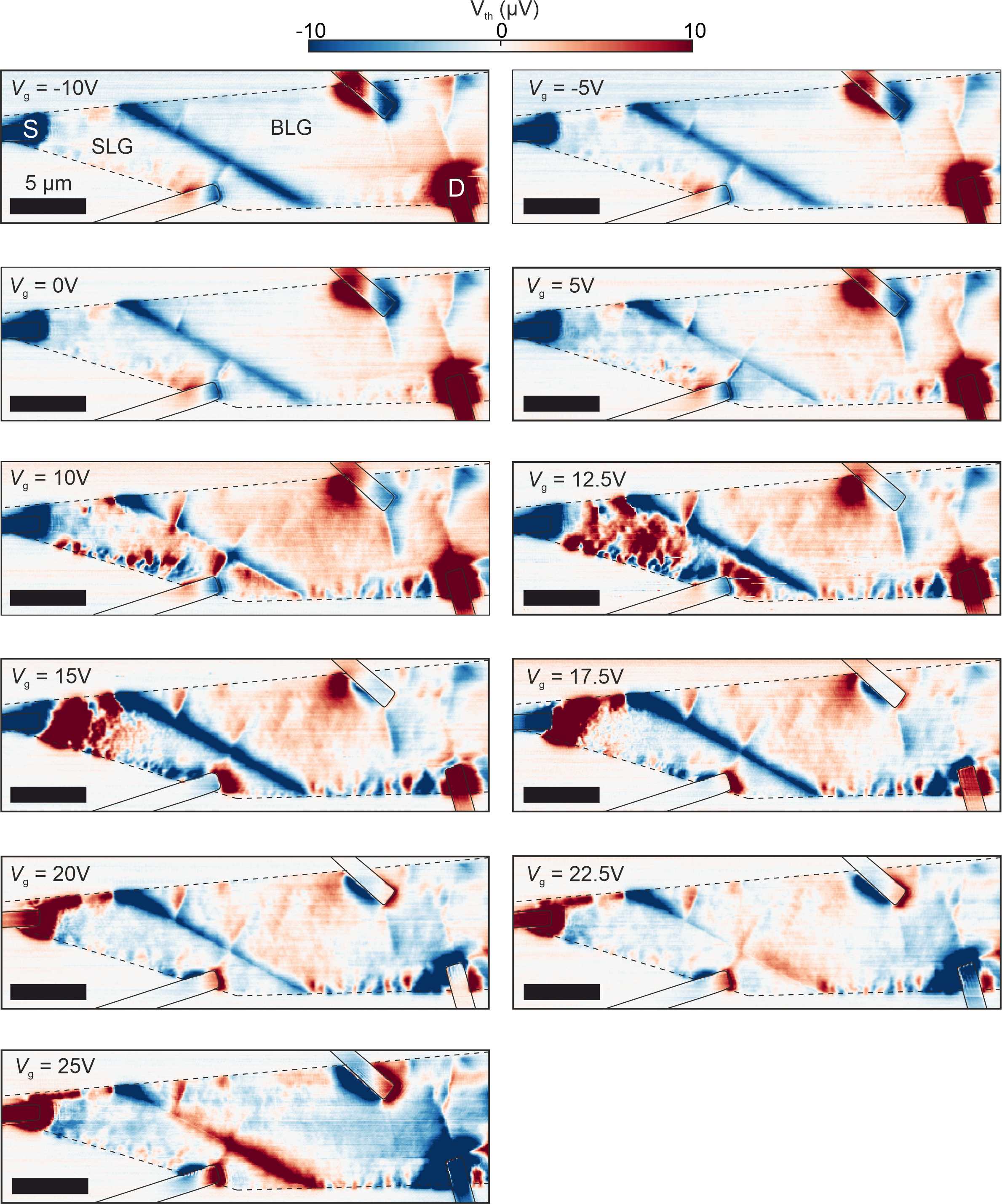}
	\caption{Additional thermovoltage maps at different gate voltages for the SLG/BLG device.}
	\label{FigSI7: Vthallgates}
\end{figure}

Additional gate dependent thermovoltage measurements of the SLG/BLG device are shown in Figure \ref{FigSI7: Vthallgates} highlighting the gate dependent effects discussed in the main text. In particular, the gate dependent strongly non-homogenous change of the carrier density due to Fermi level pinning by the contacts is clearly visible, as is the BLG gate screening. Both effects gate dependency is further shown in a GIF. 

\section{Additional simulations for a bipolar thermovoltage signal}

\begin{figure}[h!]
	\centering
	\includegraphics[width=\linewidth]{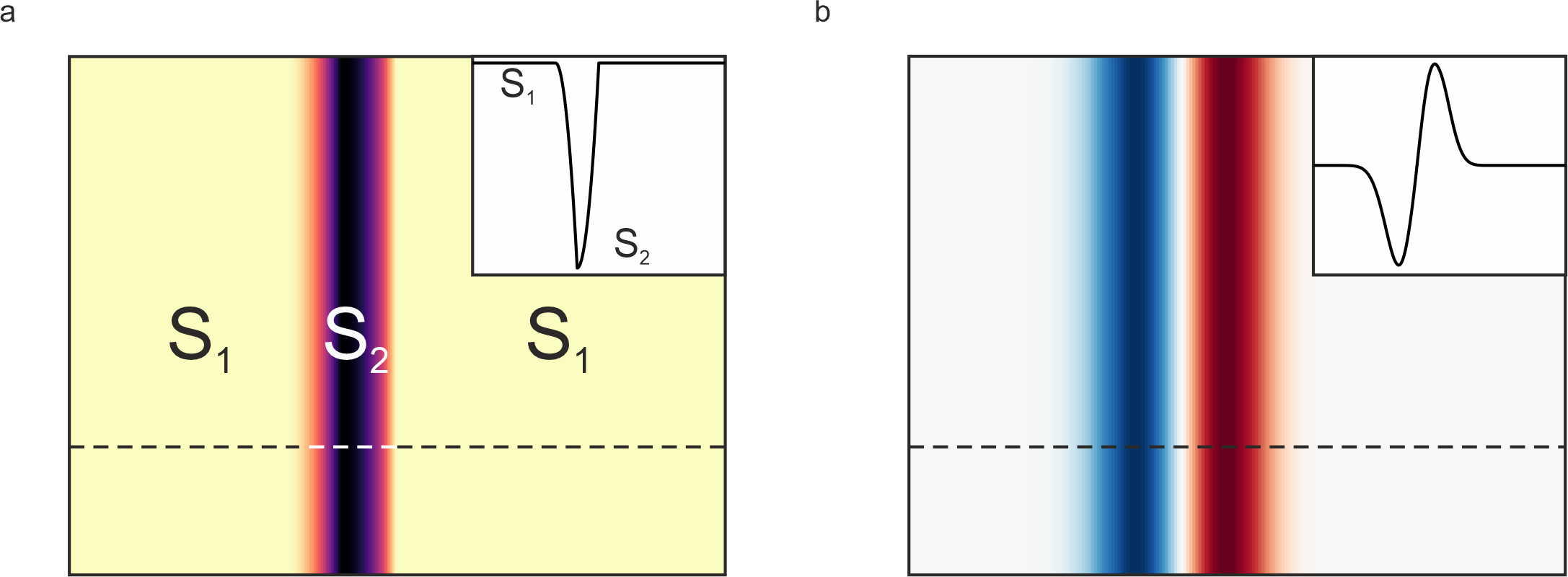}
	\caption{a) Simulated Seebeck signal dropping from $S_1$ to $S_2$ and rising back to $S_1$. The inset shows a line trace through the Seebeck map with the positon of the cut indicated by the dashed line. b) Calculated thermovoltage map. The thermovoltage map was obtained by convoluting (a) with the temperate distribution $dT(x)$. The inset shows a line cut through the thermovoltage map  with the positon of the cut indicated by the dashed line.}
	\label{FigSI6: DropsimS}
\end{figure}

In order to investigate the Seebeck coefficient distribution leading to a bipolar signal, a drop in the Seebeck coefficient is convoluted with the temperature distribution $T(x)$. The Seebeck coefficient is assumed to drop from $S_1$ to $S_2$ and rise again to $S_1$ (see Figure \ref{FigSI6: DropsimS}). This Seebeck map is then line by line 1D convoluted with the Gaussian temperature distribution to arrive at the thermovoltage map shown in \ref{FigSI6: DropsimS}b.  We note that having a differing Seebeck coefficient on either side of the drop (that is rather than rising to $S_2$ after the drop, the Seebeck coefficient rises to $S_3 \neq S_1$) does not change the bipolar nature of the resulting thermovoltage signal though it does impact the lineshape. Since the thermovoltage observed in this study is relatively symmetric we assume an approximately equal Seebeck coefficient on either side of the dip (peak) with respect to the height of the dip (peak).  

\section{Mott formula Evaluation}
In order to numerically evaluate the Mott formula in Equation (2) the measured conductance $G$ is differentiated with respect to the gate voltage $V_\mathrm{g}$. In addition, we can write the last part as
\begin{equation}
\frac{dV_g}{dE_F}(V_g) = \sqrt{\frac{|e|}{C_g \pi}}\frac{2}{\hbar v_F}\sqrt{|V_g-V_D|} ~,
\end{equation} 
where $V_D$ is the gate voltage at the Dirac point, $v_F = 10^6$ m/s is the Fermi velocity in graphene and the gate capacitance is $C_g = 115$ aF/$\upmu m^2$. The complete Mott formula used for the calculations is then 
\begin{equation}
S = -\frac{\pi^2k_B^2T}{3 |e|}\frac{1}{G}\frac{dG}{dV_g}\sqrt{\frac{|e|}{C_g \pi}}\frac{2}{\hbar v_F}\sqrt{|V_g-V_D|} ~,
\end{equation}
This can be evaluated using only the conductance trace of the device.

\section{Wrinkles in graphene}
Most of the wrinkles in graphene giving off a signal in the thermovoltage measurements can not be seen in the height/thermal conductance maps. However, some of the more prominent ones in the right side of the picture can be distingusihed in the thermal conductance map, serving as a confirmation that the bipolar thermovoltage signal is emerging from wrinkles (see Figure \ref{FigSI5: GaussFit}).

\begin{figure}[h]
	\centering
	\includegraphics[width=\linewidth]{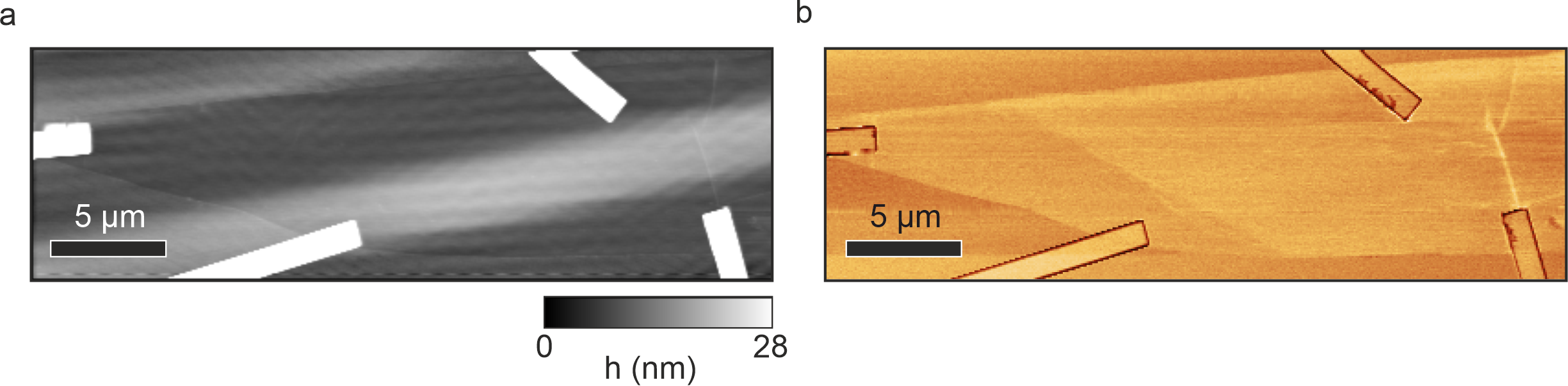}
	\caption{a) Topography map of the SLG/BLG device. b) Thermal conductance map of the SLG/BLG device.}
	\label{FigSI4: HeightThermcond}
\end{figure}

\section{Approximation of width of temperature distribution}
\begin{figure}[h]
	\centering
	\includegraphics[width=0.6\linewidth]{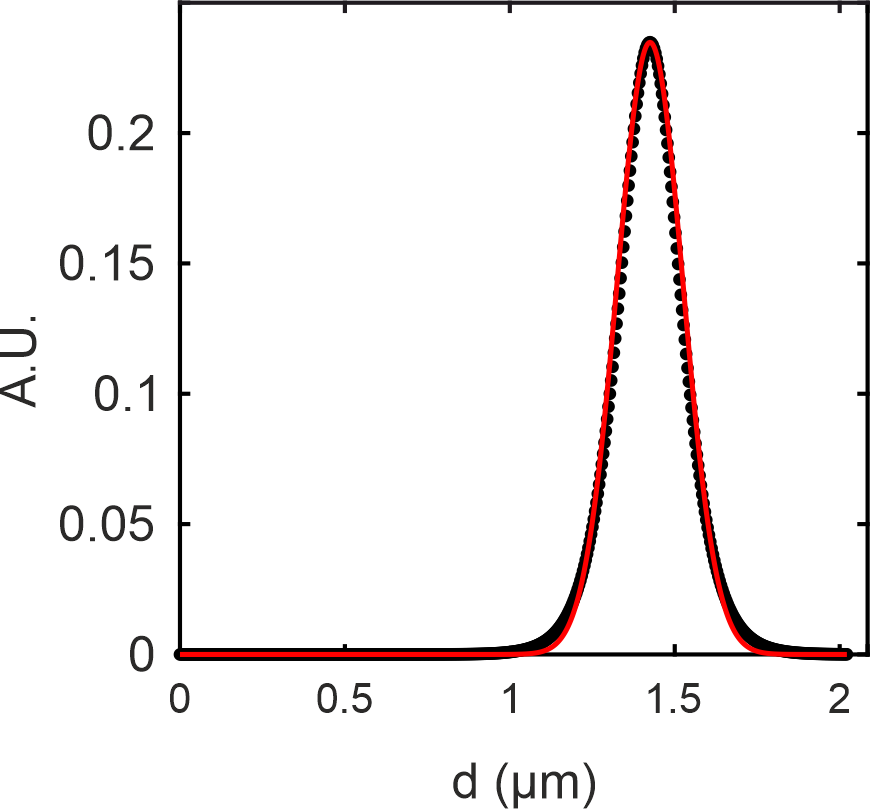}
	\caption{Fit of the derivative of the sigmoidal Boltzmann curve in Figure 3c (black dots) with a Gaussian (red line).}
	\label{FigSI5: GaussFit}
\end{figure}

In order to estimate the temperature distribution created by the tip, the derivative of the sigmoidal Boltzmann curve in Figure 3c is fitted with a Gaussian. The Gaussian has the form 
\begin{equation}
f(x) = \frac{1}{\sigma \sqrt{2\pi}}e^{\frac{-(x-\upmu)^2}{2\sigma^2}}~,
\end{equation}
where $\upmu$ is the position of the peak of the gaussian distribution and $\sigma$ is the standard deviation. The fit (see Figure \ref{FigSI5: GaussFit}) gives $\sigma = $100 nm which is then used to deconvolute the thermovoltage maps and calculate the Seebeck maps. In addition, a width of the temperature distribution of $l_\mathrm{dT}\approx 1$ $\upmu$m can be extracted.

\section{Extraction of average Thermovoltage and Seebeck coefficient from maps}

\begin{figure}[h!]
	\centering
	\includegraphics[width=\linewidth]{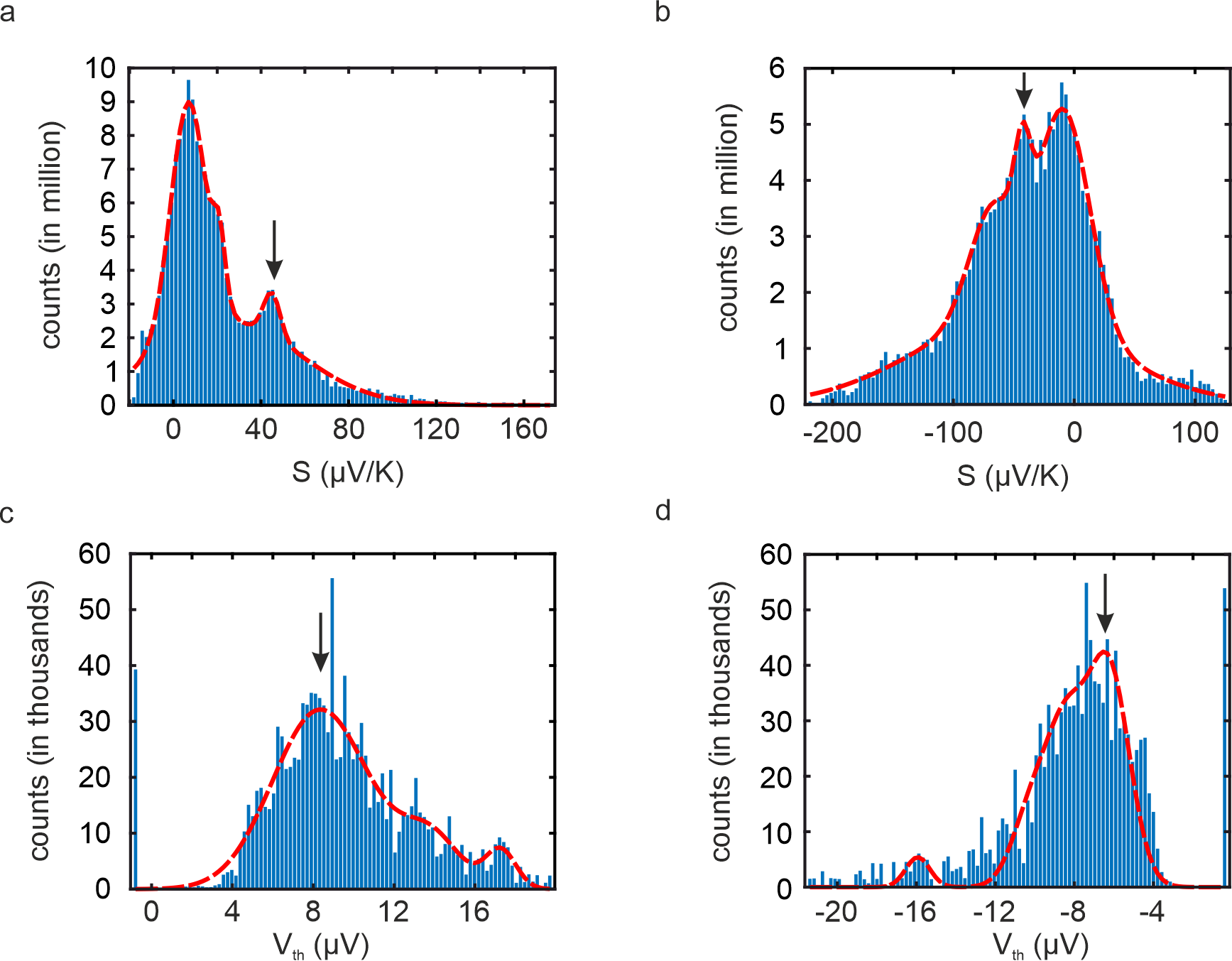}
	\caption{a-b) Bar graph showing the distribution of deconvoluted Seebeck values along the graphene channel in Figure \ref{Fig5: DeconvS}a for a p-doped (a) and n-doped (b) sample. c-d) Bar graph showing the distribution of thermovoltage values in Figure \ref{Fig2: RectDevice}a on the contacts for a p-doped (c) and n-doped (d) sample.}
	\label{FigSI5: Gaussfit}
\end{figure}

In order to extract the Seebeck coefficient of the graphene channel from the deconvoluted Seebeck maps in Figure 5a, the values in the channel are plotted in a bar graph (see Figure \ref{FigSI5: GaussFit}). We fit the distribution of values using a four peaked Gaussian to account for the multiple clusters clearly visible (see Figure \ref{FigSI5: Gaussfit}. The clusters around a Seebeck coefficient of 0 $\upmu$V/K are attributed to charge puddles in the graphene that are still noticeable at high carrier concentrations. Similarly, small clusters at exceedingly high or negative $S$ are most likely due to effects near the contact and not chosen either. The peaks used are marked by a blac 	k arrow with the error being determined by the width of the respective peak. The same procedure was repeated for the thermovoltage but here, clear outliers in the thermovoltage values around 0 $\upmu$V/K were excluded.  

\section{Signal width at $V_\mathrm{g} = 5$V}

\begin{figure}[h]
	\centering
	\includegraphics[width=0.7\linewidth]{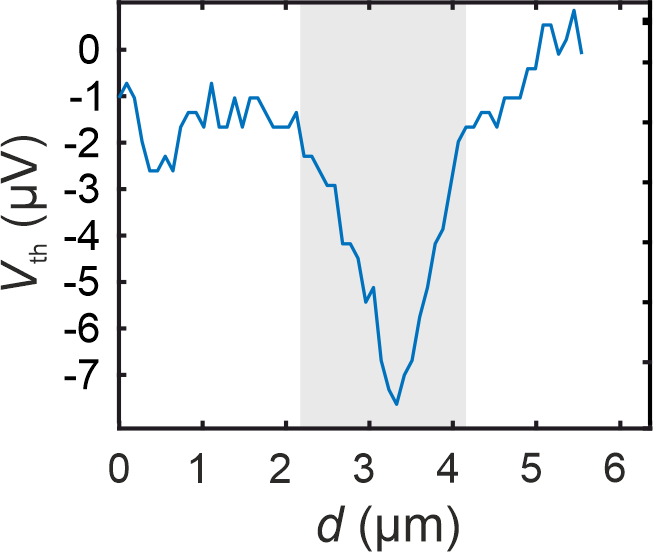}
	\caption{Linetrace through the SLG/BLG junction at $V_\mathrm{g} = 5$ V with the approximate width of the signal indicated by the gray shaded area.}
	\label{FigSI7: Width5V}
\end{figure}

As can be seen in Figure \ref{FigSI7: Width5V}, the width of the thermovoltage signal at $V_\mathrm{g} = 5$ V is approximately $l_\mathrm{d} = 2$ $\upmu$m. With $l_\mathrm{dT} \approx 1$ $\upmu$m, we can estimate $l_\mathrm{S} \approx 1$ $\upmu$m.

\begin{thebibliography}{47}
	\expandafter\ifx\csname natexlab\endcsname\relax\def\natexlab#1{#1}\fi
	\expandafter\ifx\csname bibnamefont\endcsname\relax
	\def\bibnamefont#1{#1}\fi
	\expandafter\ifx\csname bibfnamefont\endcsname\relax
	\def\bibfnamefont#1{#1}\fi
	\expandafter\ifx\csname citenamefont\endcsname\relax
	\def\citenamefont#1{#1}\fi
	\expandafter\ifx\csname url\endcsname\relax
	\def\url#1{\texttt{#1}}\fi
	\expandafter\ifx\csname urlprefix\endcsname\relax\def\urlprefix{URL }\fi
	\providecommand{\bibinfo}[2]{#2}
	\providecommand{\eprint}[2][]{\url{#2}}
	
	\bibitem[{\citenamefont{Zhang et~al.}(2014{\natexlab{a}})\citenamefont{Zhang,
			Sun, Xu, and Zhu}}]{Zhang2014}
	\bibinfo{author}{\bibfnamefont{Q.}~\bibnamefont{Zhang}},
	\bibinfo{author}{\bibfnamefont{Y.}~\bibnamefont{Sun}},
	\bibinfo{author}{\bibfnamefont{W.}~\bibnamefont{Xu}}, \bibnamefont{and}
	\bibinfo{author}{\bibfnamefont{D.}~\bibnamefont{Zhu}},
	\bibinfo{journal}{Advanced Materials} \textbf{\bibinfo{volume}{26}},
	\bibinfo{pages}{6829} (\bibinfo{year}{2014}{\natexlab{a}}), ISSN
	\bibinfo{issn}{15214095},
	\urlprefix\url{http://doi.wiley.com/10.1002/adma.201305371}.
	
	\bibitem[{\citenamefont{Crossno et~al.}(2016)\citenamefont{Crossno, Shi, Wang,
			Liu, Harzheim, Lucas, Sachdev, Kim, Taniguchi, Watanabe
			et~al.}}]{Crossno2016}
	\bibinfo{author}{\bibfnamefont{J.}~\bibnamefont{Crossno}},
	\bibinfo{author}{\bibfnamefont{J.~K.} \bibnamefont{Shi}},
	\bibinfo{author}{\bibfnamefont{K.}~\bibnamefont{Wang}},
	\bibinfo{author}{\bibfnamefont{X.}~\bibnamefont{Liu}},
	\bibinfo{author}{\bibfnamefont{A.}~\bibnamefont{Harzheim}},
	\bibinfo{author}{\bibfnamefont{A.}~\bibnamefont{Lucas}},
	\bibinfo{author}{\bibfnamefont{S.}~\bibnamefont{Sachdev}},
	\bibinfo{author}{\bibfnamefont{P.}~\bibnamefont{Kim}},
	\bibinfo{author}{\bibfnamefont{T.}~\bibnamefont{Taniguchi}},
	\bibinfo{author}{\bibfnamefont{K.}~\bibnamefont{Watanabe}},
	\bibnamefont{et~al.}, \bibinfo{journal}{Science}
	\textbf{\bibinfo{volume}{351}}, \bibinfo{pages}{1058} (\bibinfo{year}{2016}),
	\urlprefix\url{http://science.sciencemag.org.ezproxy.stanford.edu/content/351/6277/1058.abstract}.
	
	\bibitem[{\citenamefont{Harzheim
			et~al.}(2020{\natexlab{a}})\citenamefont{Harzheim, Koenemann, Gotsman,
			van~der Zant, and Gehring}}]{Harzheim2020_1}
	\bibinfo{author}{\bibfnamefont{A.}~\bibnamefont{Harzheim}},
	\bibinfo{author}{\bibfnamefont{F.}~\bibnamefont{Koenemann}},
	\bibinfo{author}{\bibfnamefont{B.}~\bibnamefont{Gotsman}},
	\bibinfo{author}{\bibfnamefont{H.}~\bibnamefont{van~der Zant}},
	\bibnamefont{and} \bibinfo{author}{\bibfnamefont{P.}~\bibnamefont{Gehring}},
	\bibinfo{journal}{In print}  (\bibinfo{year}{2020}{\natexlab{a}}).
	
	\bibitem[{\citenamefont{Kraemer et~al.}(2008)\citenamefont{Kraemer, Hu, Muto,
			Chen, Chen, and Chiesa}}]{Kraemer2008}
	\bibinfo{author}{\bibfnamefont{D.}~\bibnamefont{Kraemer}},
	\bibinfo{author}{\bibfnamefont{L.}~\bibnamefont{Hu}},
	\bibinfo{author}{\bibfnamefont{A.}~\bibnamefont{Muto}},
	\bibinfo{author}{\bibfnamefont{X.}~\bibnamefont{Chen}},
	\bibinfo{author}{\bibfnamefont{G.}~\bibnamefont{Chen}}, \bibnamefont{and}
	\bibinfo{author}{\bibfnamefont{M.}~\bibnamefont{Chiesa}},
	\bibinfo{journal}{Applied Physics Letters} \textbf{\bibinfo{volume}{92}},
	\bibinfo{pages}{243503} (\bibinfo{year}{2008}), ISSN
	\bibinfo{issn}{00036951},
	\urlprefix\url{http://aip.scitation.org/doi/10.1063/1.2947591}.
	
	\bibitem[{\citenamefont{Xia et~al.}(2009)\citenamefont{Xia, Mueller,
			Golizadeh-Mojarad, Freitage, Lin, Tsang, Perebeinos, and Avouris}}]{Xia2009}
	\bibinfo{author}{\bibfnamefont{F.}~\bibnamefont{Xia}},
	\bibinfo{author}{\bibfnamefont{T.}~\bibnamefont{Mueller}},
	\bibinfo{author}{\bibfnamefont{R.}~\bibnamefont{Golizadeh-Mojarad}},
	\bibinfo{author}{\bibfnamefont{M.}~\bibnamefont{Freitage}},
	\bibinfo{author}{\bibfnamefont{Y.~M.} \bibnamefont{Lin}},
	\bibinfo{author}{\bibfnamefont{J.}~\bibnamefont{Tsang}},
	\bibinfo{author}{\bibfnamefont{V.}~\bibnamefont{Perebeinos}},
	\bibnamefont{and} \bibinfo{author}{\bibfnamefont{P.}~\bibnamefont{Avouris}},
	\bibinfo{journal}{Nano Letters} \textbf{\bibinfo{volume}{9}},
	\bibinfo{pages}{1039} (\bibinfo{year}{2009}), ISSN \bibinfo{issn}{15306984},
	\urlprefix\url{https://pubs.acs.org/doi/10.1021/nl8033812}.
	
	\bibitem[{\citenamefont{Guo et~al.}(2018)\citenamefont{Guo, Yang, Zeng, and
			Zhang}}]{Guo2018}
	\bibinfo{author}{\bibfnamefont{S.}~\bibnamefont{Guo}},
	\bibinfo{author}{\bibfnamefont{K.}~\bibnamefont{Yang}},
	\bibinfo{author}{\bibfnamefont{Z.}~\bibnamefont{Zeng}}, \bibnamefont{and}
	\bibinfo{author}{\bibfnamefont{Y.}~\bibnamefont{Zhang}},
	\bibinfo{journal}{Physical Chemistry Chemical Physics}
	\textbf{\bibinfo{volume}{20}}, \bibinfo{pages}{14441} (\bibinfo{year}{2018}),
	ISSN \bibinfo{issn}{14639076}.
	
	\bibitem[{\citenamefont{Levander et~al.}(2011)\citenamefont{Levander, Tong, Yu,
			Suh, Fu, Zhang, Lu, Schaff, Dubon, Walukiewicz et~al.}}]{Levander2011}
	\bibinfo{author}{\bibfnamefont{A.~X.} \bibnamefont{Levander}},
	\bibinfo{author}{\bibfnamefont{T.}~\bibnamefont{Tong}},
	\bibinfo{author}{\bibfnamefont{K.~M.} \bibnamefont{Yu}},
	\bibinfo{author}{\bibfnamefont{J.}~\bibnamefont{Suh}},
	\bibinfo{author}{\bibfnamefont{D.}~\bibnamefont{Fu}},
	\bibinfo{author}{\bibfnamefont{R.}~\bibnamefont{Zhang}},
	\bibinfo{author}{\bibfnamefont{H.}~\bibnamefont{Lu}},
	\bibinfo{author}{\bibfnamefont{W.~J.} \bibnamefont{Schaff}},
	\bibinfo{author}{\bibfnamefont{O.}~\bibnamefont{Dubon}},
	\bibinfo{author}{\bibfnamefont{W.}~\bibnamefont{Walukiewicz}},
	\bibnamefont{et~al.}, \bibinfo{journal}{Applied Physics Letters}
	\textbf{\bibinfo{volume}{98}}, \bibinfo{pages}{012108}
	(\bibinfo{year}{2011}), ISSN \bibinfo{issn}{00036951},
	\urlprefix\url{http://aip.scitation.org/doi/10.1063/1.3536507}.
	
	\bibitem[{\citenamefont{Lyeo et~al.}(2004)\citenamefont{Lyeo, Khajetoorians,
			Shi, Pipe, Ram, Shakouri, and Shih}}]{Lyeo2004}
	\bibinfo{author}{\bibfnamefont{H.~K.} \bibnamefont{Lyeo}},
	\bibinfo{author}{\bibfnamefont{A.~A.} \bibnamefont{Khajetoorians}},
	\bibinfo{author}{\bibfnamefont{L.}~\bibnamefont{Shi}},
	\bibinfo{author}{\bibfnamefont{K.~P.} \bibnamefont{Pipe}},
	\bibinfo{author}{\bibfnamefont{R.~J.} \bibnamefont{Ram}},
	\bibinfo{author}{\bibfnamefont{A.}~\bibnamefont{Shakouri}}, \bibnamefont{and}
	\bibinfo{author}{\bibfnamefont{C.~K.} \bibnamefont{Shih}},
	\bibinfo{journal}{Science} \textbf{\bibinfo{volume}{303}},
	\bibinfo{pages}{816} (\bibinfo{year}{2004}), ISSN \bibinfo{issn}{00368075}.
	
	\bibitem[{\citenamefont{Zhang et~al.}(2010)\citenamefont{Zhang, Hapenciuc,
			Castillo, Borca-Tasciuc, Mehta, Karthik, and Ramanath}}]{Zhang2010}
	\bibinfo{author}{\bibfnamefont{Y.}~\bibnamefont{Zhang}},
	\bibinfo{author}{\bibfnamefont{C.~L.} \bibnamefont{Hapenciuc}},
	\bibinfo{author}{\bibfnamefont{E.~E.} \bibnamefont{Castillo}},
	\bibinfo{author}{\bibfnamefont{T.}~\bibnamefont{Borca-Tasciuc}},
	\bibinfo{author}{\bibfnamefont{R.~J.} \bibnamefont{Mehta}},
	\bibinfo{author}{\bibfnamefont{C.}~\bibnamefont{Karthik}}, \bibnamefont{and}
	\bibinfo{author}{\bibfnamefont{G.}~\bibnamefont{Ramanath}},
	\bibinfo{journal}{Applied Physics Letters} \textbf{\bibinfo{volume}{96}},
	\bibinfo{pages}{062107} (\bibinfo{year}{2010}), ISSN
	\bibinfo{issn}{00036951},
	\urlprefix\url{http://aip.scitation.org/doi/10.1063/1.3300826}.
	
	\bibitem[{\citenamefont{Sun et~al.}(2012)\citenamefont{Sun, Aivazian, Jones,
			Ross, Yao, Cobden, and Xu}}]{Sun2012}
	\bibinfo{author}{\bibfnamefont{D.}~\bibnamefont{Sun}},
	\bibinfo{author}{\bibfnamefont{G.}~\bibnamefont{Aivazian}},
	\bibinfo{author}{\bibfnamefont{A.~M.} \bibnamefont{Jones}},
	\bibinfo{author}{\bibfnamefont{J.~S.} \bibnamefont{Ross}},
	\bibinfo{author}{\bibfnamefont{W.}~\bibnamefont{Yao}},
	\bibinfo{author}{\bibfnamefont{D.}~\bibnamefont{Cobden}}, \bibnamefont{and}
	\bibinfo{author}{\bibfnamefont{X.}~\bibnamefont{Xu}},
	\bibinfo{journal}{Nature Nanotechnology} \textbf{\bibinfo{volume}{7}},
	\bibinfo{pages}{114} (\bibinfo{year}{2012}), ISSN \bibinfo{issn}{17483395}.
	
	\bibitem[{\citenamefont{Gabor et~al.}(2011)\citenamefont{Gabor, Song, Ma, Nair,
			Taychatanapat, Watanabe, Taniguchi, Levitov, and
			Jarillo-Herrero}}]{Gabor2011}
	\bibinfo{author}{\bibfnamefont{N.~M.} \bibnamefont{Gabor}},
	\bibinfo{author}{\bibfnamefont{J.~C.} \bibnamefont{Song}},
	\bibinfo{author}{\bibfnamefont{Q.}~\bibnamefont{Ma}},
	\bibinfo{author}{\bibfnamefont{N.~L.} \bibnamefont{Nair}},
	\bibinfo{author}{\bibfnamefont{T.}~\bibnamefont{Taychatanapat}},
	\bibinfo{author}{\bibfnamefont{K.}~\bibnamefont{Watanabe}},
	\bibinfo{author}{\bibfnamefont{T.}~\bibnamefont{Taniguchi}},
	\bibinfo{author}{\bibfnamefont{L.~S.} \bibnamefont{Levitov}},
	\bibnamefont{and}
	\bibinfo{author}{\bibfnamefont{P.}~\bibnamefont{Jarillo-Herrero}},
	\bibinfo{journal}{Science} \textbf{\bibinfo{volume}{334}},
	\bibinfo{pages}{648} (\bibinfo{year}{2011}), ISSN \bibinfo{issn}{10959203}.
	
	\bibitem[{\citenamefont{Zhang et~al.}(2014{\natexlab{b}})\citenamefont{Zhang,
			Yap, Li, Ng, Li, and Loh}}]{Zhang2014_1}
	\bibinfo{author}{\bibfnamefont{K.}~\bibnamefont{Zhang}},
	\bibinfo{author}{\bibfnamefont{F.~L.} \bibnamefont{Yap}},
	\bibinfo{author}{\bibfnamefont{K.}~\bibnamefont{Li}},
	\bibinfo{author}{\bibfnamefont{C.~T.} \bibnamefont{Ng}},
	\bibinfo{author}{\bibfnamefont{L.~J.} \bibnamefont{Li}}, \bibnamefont{and}
	\bibinfo{author}{\bibfnamefont{K.~P.} \bibnamefont{Loh}},
	\bibinfo{journal}{Advanced Functional Materials}
	\textbf{\bibinfo{volume}{24}}, \bibinfo{pages}{731}
	(\bibinfo{year}{2014}{\natexlab{b}}), ISSN \bibinfo{issn}{1616301X},
	\urlprefix\url{http://doi.wiley.com/10.1002/adfm.201302009}.
	
	\bibitem[{\citenamefont{Avsar et~al.}(2011)\citenamefont{Avsar, Yang, Bae,
			Balakrishnan, Volmer, Jaiswal, Yi, Ali, Gu{\"{o}}ntherodt, Hong
			et~al.}}]{Avsar2011}
	\bibinfo{author}{\bibfnamefont{A.}~\bibnamefont{Avsar}},
	\bibinfo{author}{\bibfnamefont{T.~Y.} \bibnamefont{Yang}},
	\bibinfo{author}{\bibfnamefont{S.}~\bibnamefont{Bae}},
	\bibinfo{author}{\bibfnamefont{J.}~\bibnamefont{Balakrishnan}},
	\bibinfo{author}{\bibfnamefont{F.}~\bibnamefont{Volmer}},
	\bibinfo{author}{\bibfnamefont{M.}~\bibnamefont{Jaiswal}},
	\bibinfo{author}{\bibfnamefont{Z.}~\bibnamefont{Yi}},
	\bibinfo{author}{\bibfnamefont{S.~R.} \bibnamefont{Ali}},
	\bibinfo{author}{\bibfnamefont{G.}~\bibnamefont{Gu{\"{o}}ntherodt}},
	\bibinfo{author}{\bibfnamefont{B.~H.} \bibnamefont{Hong}},
	\bibnamefont{et~al.}, \bibinfo{journal}{Nano Letters}
	\textbf{\bibinfo{volume}{11}}, \bibinfo{pages}{2363} (\bibinfo{year}{2011}),
	ISSN \bibinfo{issn}{15306984},
	\urlprefix\url{https://pubs.acs.org/doi/10.1021/nl200714q}.
	
	\bibitem[{\citenamefont{Goossens et~al.}(2017)\citenamefont{Goossens,
			Navickaite, Monasterio, Gupta, Piqueras, P{\'{e}}rez, Burwell, Nikitskiy,
			Lasanta, Gal{\'{a}}n et~al.}}]{Goossens2017}
	\bibinfo{author}{\bibfnamefont{S.}~\bibnamefont{Goossens}},
	\bibinfo{author}{\bibfnamefont{G.}~\bibnamefont{Navickaite}},
	\bibinfo{author}{\bibfnamefont{C.}~\bibnamefont{Monasterio}},
	\bibinfo{author}{\bibfnamefont{S.}~\bibnamefont{Gupta}},
	\bibinfo{author}{\bibfnamefont{J.~J.} \bibnamefont{Piqueras}},
	\bibinfo{author}{\bibfnamefont{R.}~\bibnamefont{P{\'{e}}rez}},
	\bibinfo{author}{\bibfnamefont{G.}~\bibnamefont{Burwell}},
	\bibinfo{author}{\bibfnamefont{I.}~\bibnamefont{Nikitskiy}},
	\bibinfo{author}{\bibfnamefont{T.}~\bibnamefont{Lasanta}},
	\bibinfo{author}{\bibfnamefont{T.}~\bibnamefont{Gal{\'{a}}n}},
	\bibnamefont{et~al.}, \bibinfo{journal}{Nature Photonics}
	\textbf{\bibinfo{volume}{11}}, \bibinfo{pages}{366} (\bibinfo{year}{2017}),
	ISSN \bibinfo{issn}{17494893}, \eprint{1701.03242}.
	
	\bibitem[{\citenamefont{Zuev et~al.}(2009)\citenamefont{Zuev, Chang, and
			Kim}}]{Zuev2009}
	\bibinfo{author}{\bibfnamefont{Y.~M.} \bibnamefont{Zuev}},
	\bibinfo{author}{\bibfnamefont{W.}~\bibnamefont{Chang}}, \bibnamefont{and}
	\bibinfo{author}{\bibfnamefont{P.}~\bibnamefont{Kim}},
	\bibinfo{journal}{Physical Review Letters} \textbf{\bibinfo{volume}{102}}
	(\bibinfo{year}{2009}), ISSN \bibinfo{issn}{00319007}, \eprint{0812.1393}.
	
	\bibitem[{\citenamefont{Wei et~al.}(2009)\citenamefont{Wei, Bao, Pu, Lau, and
			Shi}}]{Wei2009}
	\bibinfo{author}{\bibfnamefont{P.}~\bibnamefont{Wei}},
	\bibinfo{author}{\bibfnamefont{W.}~\bibnamefont{Bao}},
	\bibinfo{author}{\bibfnamefont{Y.}~\bibnamefont{Pu}},
	\bibinfo{author}{\bibfnamefont{C.~N.} \bibnamefont{Lau}}, \bibnamefont{and}
	\bibinfo{author}{\bibfnamefont{J.}~\bibnamefont{Shi}},
	\bibinfo{journal}{Physical Review Letters} \textbf{\bibinfo{volume}{102}}
	(\bibinfo{year}{2009}), ISSN \bibinfo{issn}{00319007}.
	
	\bibitem[{\citenamefont{Woessner et~al.}(2016)\citenamefont{Woessner,
			Alonso-Gonz{\'{a}}lez, Lundeberg, Gao, Barrios-Vargas, Navickaite, Ma,
			Janner, Watanabe, Cummings et~al.}}]{Woessner2016}
	\bibinfo{author}{\bibfnamefont{A.}~\bibnamefont{Woessner}},
	\bibinfo{author}{\bibfnamefont{P.}~\bibnamefont{Alonso-Gonz{\'{a}}lez}},
	\bibinfo{author}{\bibfnamefont{M.~B.} \bibnamefont{Lundeberg}},
	\bibinfo{author}{\bibfnamefont{Y.}~\bibnamefont{Gao}},
	\bibinfo{author}{\bibfnamefont{J.~E.} \bibnamefont{Barrios-Vargas}},
	\bibinfo{author}{\bibfnamefont{G.}~\bibnamefont{Navickaite}},
	\bibinfo{author}{\bibfnamefont{Q.}~\bibnamefont{Ma}},
	\bibinfo{author}{\bibfnamefont{D.}~\bibnamefont{Janner}},
	\bibinfo{author}{\bibfnamefont{K.}~\bibnamefont{Watanabe}},
	\bibinfo{author}{\bibfnamefont{A.~W.} \bibnamefont{Cummings}},
	\bibnamefont{et~al.}, \bibinfo{journal}{Nature Communications}
	\textbf{\bibinfo{volume}{7}} (\bibinfo{year}{2016}), ISSN
	\bibinfo{issn}{20411723}, \eprint{1508.07864}.
	
	\bibitem[{\citenamefont{Shiau et~al.}(2019)\citenamefont{Shiau, Goh, Wang, Zhu,
			Tan, Liu, and Tay}}]{Shiau2019}
	\bibinfo{author}{\bibfnamefont{L.~L.} \bibnamefont{Shiau}},
	\bibinfo{author}{\bibfnamefont{S.~C.~K.} \bibnamefont{Goh}},
	\bibinfo{author}{\bibfnamefont{X.}~\bibnamefont{Wang}},
	\bibinfo{author}{\bibfnamefont{M.}~\bibnamefont{Zhu}},
	\bibinfo{author}{\bibfnamefont{C.~S.} \bibnamefont{Tan}},
	\bibinfo{author}{\bibfnamefont{Z.}~\bibnamefont{Liu}}, \bibnamefont{and}
	\bibinfo{author}{\bibfnamefont{B.~K.} \bibnamefont{Tay}},
	\bibinfo{journal}{IEEE Transactions on Nanotechnology}
	\textbf{\bibinfo{volume}{18}}, \bibinfo{pages}{1114} (\bibinfo{year}{2019}),
	ISSN \bibinfo{issn}{19410085}.
	
	\bibitem[{\citenamefont{Harzheim et~al.}(2018)\citenamefont{Harzheim, Spiece,
			Evangeli, McCann, Falko, Sheng, Warner, Briggs, Mol, Gehring
			et~al.}}]{Harzheim2018}
	\bibinfo{author}{\bibfnamefont{A.}~\bibnamefont{Harzheim}},
	\bibinfo{author}{\bibfnamefont{J.}~\bibnamefont{Spiece}},
	\bibinfo{author}{\bibfnamefont{C.}~\bibnamefont{Evangeli}},
	\bibinfo{author}{\bibfnamefont{E.}~\bibnamefont{McCann}},
	\bibinfo{author}{\bibfnamefont{V.}~\bibnamefont{Falko}},
	\bibinfo{author}{\bibfnamefont{Y.}~\bibnamefont{Sheng}},
	\bibinfo{author}{\bibfnamefont{J.~H.} \bibnamefont{Warner}},
	\bibinfo{author}{\bibfnamefont{G.~A.~D.} \bibnamefont{Briggs}},
	\bibinfo{author}{\bibfnamefont{J.~A.} \bibnamefont{Mol}},
	\bibinfo{author}{\bibfnamefont{P.}~\bibnamefont{Gehring}},
	\bibnamefont{et~al.}, \bibinfo{journal}{Nano Letters}
	\textbf{\bibinfo{volume}{18}}, \bibinfo{pages}{7719} (\bibinfo{year}{2018}),
	ISSN \bibinfo{issn}{15306992},
	\urlprefix\url{https://pubs.acs.org/doi/10.1021/acs.nanolett.8b03406}.
	
	\bibitem[{\citenamefont{Gom{\`{e}}s et~al.}(2015)\citenamefont{Gom{\`{e}}s,
			Assy, and Chapuis}}]{Gomes2015}
	\bibinfo{author}{\bibfnamefont{S.}~\bibnamefont{Gom{\`{e}}s}},
	\bibinfo{author}{\bibfnamefont{A.}~\bibnamefont{Assy}}, \bibnamefont{and}
	\bibinfo{author}{\bibfnamefont{P.~O.} \bibnamefont{Chapuis}},
	\emph{\bibinfo{title}{{Scanning thermal microscopy: A review}}}
	(\bibinfo{year}{2015}),
	\urlprefix\url{http://doi.wiley.com/10.1002/pssa.201400360}.
	
	\bibitem[{\citenamefont{Xu et~al.}(2010)\citenamefont{Xu, Gabor, Alden, {Van
				Der Zande}, and McEuen}}]{Xu2010}
	\bibinfo{author}{\bibfnamefont{X.}~\bibnamefont{Xu}},
	\bibinfo{author}{\bibfnamefont{N.~M.} \bibnamefont{Gabor}},
	\bibinfo{author}{\bibfnamefont{J.~S.} \bibnamefont{Alden}},
	\bibinfo{author}{\bibfnamefont{A.~M.} \bibnamefont{{Van Der Zande}}},
	\bibnamefont{and} \bibinfo{author}{\bibfnamefont{P.~L.}
		\bibnamefont{McEuen}}, \bibinfo{journal}{Nano Letters}
	\textbf{\bibinfo{volume}{10}}, \bibinfo{pages}{562} (\bibinfo{year}{2010}),
	ISSN \bibinfo{issn}{15306984}, \eprint{0907.3173},
	\urlprefix\url{https://pubs.acs.org/doi/pdf/10.1021/nl903451y}.
	
	\bibitem[{\citenamefont{Hach{\'{e}} et~al.}(2012)\citenamefont{Hach{\'{e}}, Do,
			and Bonora}}]{Hache2012}
	\bibinfo{author}{\bibfnamefont{A.}~\bibnamefont{Hach{\'{e}}}},
	\bibinfo{author}{\bibfnamefont{P.~A.} \bibnamefont{Do}}, \bibnamefont{and}
	\bibinfo{author}{\bibfnamefont{S.}~\bibnamefont{Bonora}},
	\bibinfo{journal}{Applied Optics} \textbf{\bibinfo{volume}{51}},
	\bibinfo{pages}{6578} (\bibinfo{year}{2012}), ISSN \bibinfo{issn}{15394522}.
	
	\bibitem[{\citenamefont{Shi et~al.}(2009)\citenamefont{Shi, Dong, Chen, Wang,
			and Li}}]{Shi2009}
	\bibinfo{author}{\bibfnamefont{Y.}~\bibnamefont{Shi}},
	\bibinfo{author}{\bibfnamefont{X.}~\bibnamefont{Dong}},
	\bibinfo{author}{\bibfnamefont{P.}~\bibnamefont{Chen}},
	\bibinfo{author}{\bibfnamefont{J.}~\bibnamefont{Wang}}, \bibnamefont{and}
	\bibinfo{author}{\bibfnamefont{L.~J.} \bibnamefont{Li}},
	\bibinfo{journal}{Physical Review B - Condensed Matter and Materials Physics}
	\textbf{\bibinfo{volume}{79}} (\bibinfo{year}{2009}), ISSN
	\bibinfo{issn}{10980121}.
	
	\bibitem[{\citenamefont{Wang et~al.}(2012)\citenamefont{Wang, Jin, Kim, Hilmer,
			Paulus, Shih, Ham, Sanchez-Yamagishi, Watanabe, Taniguchi et~al.}}]{Wang2012}
	\bibinfo{author}{\bibfnamefont{Q.~H.} \bibnamefont{Wang}},
	\bibinfo{author}{\bibfnamefont{Z.}~\bibnamefont{Jin}},
	\bibinfo{author}{\bibfnamefont{K.~K.} \bibnamefont{Kim}},
	\bibinfo{author}{\bibfnamefont{A.~J.} \bibnamefont{Hilmer}},
	\bibinfo{author}{\bibfnamefont{G.~L.} \bibnamefont{Paulus}},
	\bibinfo{author}{\bibfnamefont{C.~J.} \bibnamefont{Shih}},
	\bibinfo{author}{\bibfnamefont{M.~H.} \bibnamefont{Ham}},
	\bibinfo{author}{\bibfnamefont{J.~D.} \bibnamefont{Sanchez-Yamagishi}},
	\bibinfo{author}{\bibfnamefont{K.}~\bibnamefont{Watanabe}},
	\bibinfo{author}{\bibfnamefont{T.}~\bibnamefont{Taniguchi}},
	\bibnamefont{et~al.}, \bibinfo{journal}{Nature Chemistry}
	\textbf{\bibinfo{volume}{4}}, \bibinfo{pages}{724} (\bibinfo{year}{2012}),
	ISSN \bibinfo{issn}{17554330}.
	
	\bibitem[{\citenamefont{Xue et~al.}(2011)\citenamefont{Xue, Sanchez-Yamagishi,
			Bulmash, Jacquod, Deshpande, Watanabe, Taniguchi, Jarillo-Herrero, and
			Leroy}}]{Xue2011}
	\bibinfo{author}{\bibfnamefont{J.}~\bibnamefont{Xue}},
	\bibinfo{author}{\bibfnamefont{J.}~\bibnamefont{Sanchez-Yamagishi}},
	\bibinfo{author}{\bibfnamefont{D.}~\bibnamefont{Bulmash}},
	\bibinfo{author}{\bibfnamefont{P.}~\bibnamefont{Jacquod}},
	\bibinfo{author}{\bibfnamefont{A.}~\bibnamefont{Deshpande}},
	\bibinfo{author}{\bibfnamefont{K.}~\bibnamefont{Watanabe}},
	\bibinfo{author}{\bibfnamefont{T.}~\bibnamefont{Taniguchi}},
	\bibinfo{author}{\bibfnamefont{P.}~\bibnamefont{Jarillo-Herrero}},
	\bibnamefont{and} \bibinfo{author}{\bibfnamefont{B.~J.} \bibnamefont{Leroy}},
	\bibinfo{journal}{Nature Materials} \textbf{\bibinfo{volume}{10}},
	\bibinfo{pages}{282} (\bibinfo{year}{2011}), ISSN \bibinfo{issn}{14764660},
	\eprint{1102.2642}.
	
	\bibitem[{\citenamefont{Tovee et~al.}(2012)\citenamefont{Tovee, Pumarol, Zeze,
			Kjoller, and Kolosov}}]{Tovee2012}
	\bibinfo{author}{\bibfnamefont{P.}~\bibnamefont{Tovee}},
	\bibinfo{author}{\bibfnamefont{M.}~\bibnamefont{Pumarol}},
	\bibinfo{author}{\bibfnamefont{D.}~\bibnamefont{Zeze}},
	\bibinfo{author}{\bibfnamefont{K.}~\bibnamefont{Kjoller}}, \bibnamefont{and}
	\bibinfo{author}{\bibfnamefont{O.}~\bibnamefont{Kolosov}},
	\bibinfo{journal}{Journal of Applied Physics} \textbf{\bibinfo{volume}{112}},
	\bibinfo{pages}{114317} (\bibinfo{year}{2012}), ISSN
	\bibinfo{issn}{00218979}, \eprint{1110.6055},
	\urlprefix\url{http://aip.scitation.org/doi/10.1063/1.4767923}.
	
	\bibitem[{\citenamefont{Lee et~al.}(2008)\citenamefont{Lee, Balasubramanian,
			Weitz, Burghard, and Kern}}]{Lee2008}
	\bibinfo{author}{\bibfnamefont{E.~J.} \bibnamefont{Lee}},
	\bibinfo{author}{\bibfnamefont{K.}~\bibnamefont{Balasubramanian}},
	\bibinfo{author}{\bibfnamefont{R.~T.} \bibnamefont{Weitz}},
	\bibinfo{author}{\bibfnamefont{M.}~\bibnamefont{Burghard}}, \bibnamefont{and}
	\bibinfo{author}{\bibfnamefont{K.}~\bibnamefont{Kern}},
	\bibinfo{journal}{Nature Nanotechnology} \textbf{\bibinfo{volume}{3}},
	\bibinfo{pages}{486} (\bibinfo{year}{2008}), ISSN \bibinfo{issn}{17483395}.
	
	\bibitem[{\citenamefont{Sun et~al.}(2010)\citenamefont{Sun, Divin, Berger, {De
				Heer}, First, and Norris}}]{Sun2010}
	\bibinfo{author}{\bibfnamefont{D.}~\bibnamefont{Sun}},
	\bibinfo{author}{\bibfnamefont{C.}~\bibnamefont{Divin}},
	\bibinfo{author}{\bibfnamefont{C.}~\bibnamefont{Berger}},
	\bibinfo{author}{\bibfnamefont{W.~A.} \bibnamefont{{De Heer}}},
	\bibinfo{author}{\bibfnamefont{P.~N.} \bibnamefont{First}}, \bibnamefont{and}
	\bibinfo{author}{\bibfnamefont{T.~B.} \bibnamefont{Norris}},
	\bibinfo{journal}{Physical Review Letters} \textbf{\bibinfo{volume}{104}}
	(\bibinfo{year}{2010}), ISSN \bibinfo{issn}{00319007}.
	
	\bibitem[{\citenamefont{Ohta et~al.}(2007)\citenamefont{Ohta, Kim, Mune,
			Mizoguchi, Nomura, Ohta, Nomura, Nakanishi, Ikuhara, Hirano
			et~al.}}]{Ohta2007}
	\bibinfo{author}{\bibfnamefont{H.}~\bibnamefont{Ohta}},
	\bibinfo{author}{\bibfnamefont{S.}~\bibnamefont{Kim}},
	\bibinfo{author}{\bibfnamefont{Y.}~\bibnamefont{Mune}},
	\bibinfo{author}{\bibfnamefont{T.}~\bibnamefont{Mizoguchi}},
	\bibinfo{author}{\bibfnamefont{K.}~\bibnamefont{Nomura}},
	\bibinfo{author}{\bibfnamefont{S.}~\bibnamefont{Ohta}},
	\bibinfo{author}{\bibfnamefont{T.}~\bibnamefont{Nomura}},
	\bibinfo{author}{\bibfnamefont{Y.}~\bibnamefont{Nakanishi}},
	\bibinfo{author}{\bibfnamefont{Y.}~\bibnamefont{Ikuhara}},
	\bibinfo{author}{\bibfnamefont{M.}~\bibnamefont{Hirano}},
	\bibnamefont{et~al.}, \bibinfo{journal}{Nature Materials}
	\textbf{\bibinfo{volume}{6}}, \bibinfo{pages}{129} (\bibinfo{year}{2007}),
	ISSN \bibinfo{issn}{14764660}.
	
	\bibitem[{\citenamefont{Rojo et~al.}(2019)\citenamefont{Rojo, Li, Sievers,
			Bornstein, Yalon, Deshmukh, Vaziri, Bae, Xiong, Donadio et~al.}}]{Rojo2019}
	\bibinfo{author}{\bibfnamefont{M.~M.} \bibnamefont{Rojo}},
	\bibinfo{author}{\bibfnamefont{Z.}~\bibnamefont{Li}},
	\bibinfo{author}{\bibfnamefont{C.}~\bibnamefont{Sievers}},
	\bibinfo{author}{\bibfnamefont{A.~C.} \bibnamefont{Bornstein}},
	\bibinfo{author}{\bibfnamefont{E.}~\bibnamefont{Yalon}},
	\bibinfo{author}{\bibfnamefont{S.}~\bibnamefont{Deshmukh}},
	\bibinfo{author}{\bibfnamefont{S.}~\bibnamefont{Vaziri}},
	\bibinfo{author}{\bibfnamefont{M.~H.} \bibnamefont{Bae}},
	\bibinfo{author}{\bibfnamefont{F.}~\bibnamefont{Xiong}},
	\bibinfo{author}{\bibfnamefont{D.}~\bibnamefont{Donadio}},
	\bibnamefont{et~al.}, \bibinfo{journal}{2D Materials}
	\textbf{\bibinfo{volume}{6}}, \bibinfo{pages}{011005} (\bibinfo{year}{2019}),
	ISSN \bibinfo{issn}{20531583}.
	
	\bibitem[{\citenamefont{Yoon et~al.}(2011)\citenamefont{Yoon, Son, and
			Cheong}}]{Yoon2011}
	\bibinfo{author}{\bibfnamefont{D.}~\bibnamefont{Yoon}},
	\bibinfo{author}{\bibfnamefont{Y.~W.} \bibnamefont{Son}}, \bibnamefont{and}
	\bibinfo{author}{\bibfnamefont{H.}~\bibnamefont{Cheong}},
	\bibinfo{journal}{Physical Review Letters} \textbf{\bibinfo{volume}{106}}
	(\bibinfo{year}{2011}), ISSN \bibinfo{issn}{00319007}, \eprint{1103.3147}.
	
	\bibitem[{\citenamefont{Lu et~al.}(2020)\citenamefont{Lu, Chu, and
			An}}]{Lu2020}
	\bibinfo{author}{\bibfnamefont{H.}~\bibnamefont{Lu}},
	\bibinfo{author}{\bibfnamefont{P.~K.} \bibnamefont{Chu}}, \bibnamefont{and}
	\bibinfo{author}{\bibfnamefont{Z.}~\bibnamefont{An}},
	\bibinfo{journal}{Small} p. \bibinfo{pages}{1907170} (\bibinfo{year}{2020}),
	ISSN \bibinfo{issn}{1613-6810},
	\urlprefix\url{https://onlinelibrary.wiley.com/doi/abs/10.1002/smll.201907170}.
	
	\bibitem[{\citenamefont{Mueller et~al.}(2009)\citenamefont{Mueller, Xia,
			Freitag, Tsang, and Avouris}}]{Mueller2009}
	\bibinfo{author}{\bibfnamefont{T.}~\bibnamefont{Mueller}},
	\bibinfo{author}{\bibfnamefont{F.}~\bibnamefont{Xia}},
	\bibinfo{author}{\bibfnamefont{M.}~\bibnamefont{Freitag}},
	\bibinfo{author}{\bibfnamefont{J.}~\bibnamefont{Tsang}}, \bibnamefont{and}
	\bibinfo{author}{\bibfnamefont{P.}~\bibnamefont{Avouris}},
	\bibinfo{journal}{Physical Review B - Condensed Matter and Materials Physics}
	\textbf{\bibinfo{volume}{79}} (\bibinfo{year}{2009}), ISSN
	\bibinfo{issn}{10980121}.
	
	\bibitem[{\citenamefont{McCann and Koshino}(2013)}]{McCann2013}
	\bibinfo{author}{\bibfnamefont{E.}~\bibnamefont{McCann}} \bibnamefont{and}
	\bibinfo{author}{\bibfnamefont{M.}~\bibnamefont{Koshino}},
	\bibinfo{journal}{Reports on Progress in Physics}
	\textbf{\bibinfo{volume}{76}} (\bibinfo{year}{2013}), ISSN
	\bibinfo{issn}{00344885}, \eprint{1205.6953}.
	
	\bibitem[{\citenamefont{{Castro Neto} et~al.}(2009)\citenamefont{{Castro Neto},
			Guinea, Peres, Novoselov, and Geim}}]{CastroNeto2009}
	\bibinfo{author}{\bibfnamefont{A.~H.} \bibnamefont{{Castro Neto}}},
	\bibinfo{author}{\bibfnamefont{F.}~\bibnamefont{Guinea}},
	\bibinfo{author}{\bibfnamefont{N.~M.} \bibnamefont{Peres}},
	\bibinfo{author}{\bibfnamefont{K.~S.} \bibnamefont{Novoselov}},
	\bibnamefont{and} \bibinfo{author}{\bibfnamefont{A.~K.} \bibnamefont{Geim}},
	\bibinfo{journal}{Reviews of Modern Physics} \textbf{\bibinfo{volume}{81}},
	\bibinfo{pages}{109} (\bibinfo{year}{2009}), ISSN \bibinfo{issn}{00346861},
	\eprint{0709.1163}.
	
	\bibitem[{\citenamefont{Nguyen et~al.}(2015)\citenamefont{Nguyen, Nguyen,
			Nguyen, Saint-Martin, and Dollfus}}]{Nguyen2015}
	\bibinfo{author}{\bibfnamefont{M.~C.} \bibnamefont{Nguyen}},
	\bibinfo{author}{\bibfnamefont{V.~H.} \bibnamefont{Nguyen}},
	\bibinfo{author}{\bibfnamefont{H.~V.} \bibnamefont{Nguyen}},
	\bibinfo{author}{\bibfnamefont{J.}~\bibnamefont{Saint-Martin}},
	\bibnamefont{and} \bibinfo{author}{\bibfnamefont{P.}~\bibnamefont{Dollfus}},
	\bibinfo{journal}{Physica E: Low-Dimensional Systems and Nanostructures}
	\textbf{\bibinfo{volume}{73}}, \bibinfo{pages}{207} (\bibinfo{year}{2015}),
	ISSN \bibinfo{issn}{13869477}, \eprint{1505.06474}.
	
	\bibitem[{\citenamefont{Tovee et~al.}(2014)\citenamefont{Tovee, Pumarol,
			Rosamond, Jones, Petty, Zeze, and Kolosov}}]{Tovee2014}
	\bibinfo{author}{\bibfnamefont{P.~D.} \bibnamefont{Tovee}},
	\bibinfo{author}{\bibfnamefont{M.~E.} \bibnamefont{Pumarol}},
	\bibinfo{author}{\bibfnamefont{M.~C.} \bibnamefont{Rosamond}},
	\bibinfo{author}{\bibfnamefont{R.}~\bibnamefont{Jones}},
	\bibinfo{author}{\bibfnamefont{M.~C.} \bibnamefont{Petty}},
	\bibinfo{author}{\bibfnamefont{D.~A.} \bibnamefont{Zeze}}, \bibnamefont{and}
	\bibinfo{author}{\bibfnamefont{O.~V.} \bibnamefont{Kolosov}},
	\bibinfo{journal}{Physical Chemistry Chemical Physics}
	\textbf{\bibinfo{volume}{16}}, \bibinfo{pages}{1174} (\bibinfo{year}{2014}),
	ISSN \bibinfo{issn}{14639076}.
	
	\bibitem[{\citenamefont{Vera-Marun et~al.}(2016)\citenamefont{Vera-Marun,
			van~den Berg, Dejene, and van Wees}}]{Vera-Marun2016}
	\bibinfo{author}{\bibfnamefont{I.~J.} \bibnamefont{Vera-Marun}},
	\bibinfo{author}{\bibfnamefont{J.~J.} \bibnamefont{van~den Berg}},
	\bibinfo{author}{\bibfnamefont{F.~K.} \bibnamefont{Dejene}},
	\bibnamefont{and} \bibinfo{author}{\bibfnamefont{B.~J.} \bibnamefont{van
			Wees}}, \bibinfo{journal}{Nature Communications}
	\textbf{\bibinfo{volume}{7}}, \bibinfo{pages}{11525} (\bibinfo{year}{2016}),
	ISSN \bibinfo{issn}{2041-1723},
	\urlprefix\url{http://www.nature.com/doifinder/10.1038/ncomms11525}.
	
	\bibitem[{\citenamefont{Harzheim
			et~al.}(2020{\natexlab{b}})\citenamefont{Harzheim, Sowa, Swett, Briggs, Mol,
			and Gehring}}]{Harzheim2020}
	\bibinfo{author}{\bibfnamefont{A.}~\bibnamefont{Harzheim}},
	\bibinfo{author}{\bibfnamefont{J.~K.} \bibnamefont{Sowa}},
	\bibinfo{author}{\bibfnamefont{J.~L.} \bibnamefont{Swett}},
	\bibinfo{author}{\bibfnamefont{G.~A.~D.} \bibnamefont{Briggs}},
	\bibinfo{author}{\bibfnamefont{J.~A.} \bibnamefont{Mol}}, \bibnamefont{and}
	\bibinfo{author}{\bibfnamefont{P.}~\bibnamefont{Gehring}},
	\bibinfo{journal}{Physical Review Research} \textbf{\bibinfo{volume}{2}},
	\bibinfo{pages}{013140} (\bibinfo{year}{2020}{\natexlab{b}}),
	\eprint{1906.05401}, \urlprefix\url{http://arxiv.org/abs/1906.05401}.
	
	\bibitem[{\citenamefont{Gehring et~al.}(2019)\citenamefont{Gehring, {Van Der
				Star}, Evangeli, {Le Roy}, Bogani, Kolosov, and {Van Der
				Zant}}}]{Gehring2019}
	\bibinfo{author}{\bibfnamefont{P.}~\bibnamefont{Gehring}},
	\bibinfo{author}{\bibfnamefont{M.}~\bibnamefont{{Van Der Star}}},
	\bibinfo{author}{\bibfnamefont{C.}~\bibnamefont{Evangeli}},
	\bibinfo{author}{\bibfnamefont{J.~J.} \bibnamefont{{Le Roy}}},
	\bibinfo{author}{\bibfnamefont{L.}~\bibnamefont{Bogani}},
	\bibinfo{author}{\bibfnamefont{O.~V.} \bibnamefont{Kolosov}},
	\bibnamefont{and} \bibinfo{author}{\bibfnamefont{H.~S.} \bibnamefont{{Van Der
				Zant}}}, \bibinfo{journal}{Applied Physics Letters}
	\textbf{\bibinfo{volume}{115}} (\bibinfo{year}{2019}), ISSN
	\bibinfo{issn}{00036951}, \eprint{1907.02815}.
	
	\bibitem[{\citenamefont{Gehring et~al.}(2017)\citenamefont{Gehring, Harzheim,
			Spi{\`{e}}ce, Sheng, Rogers, Evangeli, Mishra, Robinson, Porfyrakis, Warner
			et~al.}}]{Gehring2017}
	\bibinfo{author}{\bibfnamefont{P.}~\bibnamefont{Gehring}},
	\bibinfo{author}{\bibfnamefont{A.}~\bibnamefont{Harzheim}},
	\bibinfo{author}{\bibfnamefont{J.}~\bibnamefont{Spi{\`{e}}ce}},
	\bibinfo{author}{\bibfnamefont{Y.}~\bibnamefont{Sheng}},
	\bibinfo{author}{\bibfnamefont{G.}~\bibnamefont{Rogers}},
	\bibinfo{author}{\bibfnamefont{C.}~\bibnamefont{Evangeli}},
	\bibinfo{author}{\bibfnamefont{A.}~\bibnamefont{Mishra}},
	\bibinfo{author}{\bibfnamefont{B.~J.} \bibnamefont{Robinson}},
	\bibinfo{author}{\bibfnamefont{K.}~\bibnamefont{Porfyrakis}},
	\bibinfo{author}{\bibfnamefont{J.~H.} \bibnamefont{Warner}},
	\bibnamefont{et~al.}, \bibinfo{journal}{Nano Letters}
	\textbf{\bibinfo{volume}{17}}, \bibinfo{pages}{7055} (\bibinfo{year}{2017}),
	ISSN \bibinfo{issn}{15306992}, \eprint{1710.08344},
	\urlprefix\url{http://pubs.acs.org/doi/abs/10.1021/acs.nanolett.7b03736}.
	
	\bibitem[{\citenamefont{Menges et~al.}(2016)\citenamefont{Menges, Mensch,
			Schmid, Riel, Stemmer, and Gotsmann}}]{Menges2016}
	\bibinfo{author}{\bibfnamefont{F.}~\bibnamefont{Menges}},
	\bibinfo{author}{\bibfnamefont{P.}~\bibnamefont{Mensch}},
	\bibinfo{author}{\bibfnamefont{H.}~\bibnamefont{Schmid}},
	\bibinfo{author}{\bibfnamefont{H.}~\bibnamefont{Riel}},
	\bibinfo{author}{\bibfnamefont{A.}~\bibnamefont{Stemmer}}, \bibnamefont{and}
	\bibinfo{author}{\bibfnamefont{B.}~\bibnamefont{Gotsmann}},
	\bibinfo{journal}{Nature Communications} \textbf{\bibinfo{volume}{7}},
	\bibinfo{pages}{10874} (\bibinfo{year}{2016}), ISSN \bibinfo{issn}{20411723},
	\urlprefix\url{http://www.nature.com/doifinder/10.1038/ncomms10874}.
	
	\bibitem[{\citenamefont{Kim et~al.}(2012)\citenamefont{Kim, Jeong, Lee, and
			Reddy}}]{Kim2012}
	\bibinfo{author}{\bibfnamefont{K.}~\bibnamefont{Kim}},
	\bibinfo{author}{\bibfnamefont{W.}~\bibnamefont{Jeong}},
	\bibinfo{author}{\bibfnamefont{W.}~\bibnamefont{Lee}}, \bibnamefont{and}
	\bibinfo{author}{\bibfnamefont{P.}~\bibnamefont{Reddy}},
	\bibinfo{journal}{ACS Nano} \textbf{\bibinfo{volume}{6}},
	\bibinfo{pages}{4248} (\bibinfo{year}{2012}), ISSN \bibinfo{issn}{19360851},
	\urlprefix\url{www.acsnano.org}.
	
	\bibitem[{\citenamefont{Lee et~al.}(2012)\citenamefont{Lee, Kim, Lee, Kim, Lim,
			Kwon, and Lee}}]{Lee2012}
	\bibinfo{author}{\bibfnamefont{B.}~\bibnamefont{Lee}},
	\bibinfo{author}{\bibfnamefont{K.}~\bibnamefont{Kim}},
	\bibinfo{author}{\bibfnamefont{S.}~\bibnamefont{Lee}},
	\bibinfo{author}{\bibfnamefont{J.~H.} \bibnamefont{Kim}},
	\bibinfo{author}{\bibfnamefont{D.~S.} \bibnamefont{Lim}},
	\bibinfo{author}{\bibfnamefont{O.}~\bibnamefont{Kwon}}, \bibnamefont{and}
	\bibinfo{author}{\bibfnamefont{J.~S.} \bibnamefont{Lee}},
	\bibinfo{journal}{Nano Letters} \textbf{\bibinfo{volume}{12}},
	\bibinfo{pages}{4472} (\bibinfo{year}{2012}), ISSN \bibinfo{issn}{15306984},
	\urlprefix\url{http://pubs.acs.org/doi/10.1021/nl301359c}.
	
	\bibitem[{\citenamefont{Evangeli et~al.}(2019)\citenamefont{Evangeli, Spiece,
			Sangtarash, Molina-Mendoza, Mucientes, Mueller, Lambert, Sadeghi, and
			Kolosov}}]{Evangeli2019}
	\bibinfo{author}{\bibfnamefont{C.}~\bibnamefont{Evangeli}},
	\bibinfo{author}{\bibfnamefont{J.}~\bibnamefont{Spiece}},
	\bibinfo{author}{\bibfnamefont{S.}~\bibnamefont{Sangtarash}},
	\bibinfo{author}{\bibfnamefont{A.~J.} \bibnamefont{Molina-Mendoza}},
	\bibinfo{author}{\bibfnamefont{M.}~\bibnamefont{Mucientes}},
	\bibinfo{author}{\bibfnamefont{T.}~\bibnamefont{Mueller}},
	\bibinfo{author}{\bibfnamefont{C.}~\bibnamefont{Lambert}},
	\bibinfo{author}{\bibfnamefont{H.}~\bibnamefont{Sadeghi}}, \bibnamefont{and}
	\bibinfo{author}{\bibfnamefont{O.}~\bibnamefont{Kolosov}},
	\bibinfo{journal}{Advanced Electronic Materials}
	\textbf{\bibinfo{volume}{5}}, \bibinfo{pages}{1900331}
	(\bibinfo{year}{2019}), ISSN \bibinfo{issn}{2199160X},
	\urlprefix\url{https://onlinelibrary.wiley.com/doi/abs/10.1002/aelm.201900331}.
	
	\bibitem[{\citenamefont{Hunt}(1973)}]{Hunt1973}
	\bibinfo{author}{\bibfnamefont{B.~R.} \bibnamefont{Hunt}},
	\bibinfo{journal}{IEEE Transactions on Computers}
	\textbf{\bibinfo{volume}{C-22}}, \bibinfo{pages}{805} (\bibinfo{year}{1973}),
	ISSN \bibinfo{issn}{00189340},
	\urlprefix\url{https://ieeexplore.ieee.org/stamp/stamp.jsp?tp={\&}arnumber=5009169}.
	
	\bibitem[{\citenamefont{Lagendijk and Biemond}(2009)}]{Lagendijk2009}
	\bibinfo{author}{\bibfnamefont{R.~L.} \bibnamefont{Lagendijk}}
	\bibnamefont{and} \bibinfo{author}{\bibfnamefont{J.}~\bibnamefont{Biemond}},
	in \emph{\bibinfo{booktitle}{The Essential Guide to Image Processing}}
	(\bibinfo{publisher}{Elsevier Inc.}, \bibinfo{year}{2009}), pp.
	\bibinfo{pages}{323--348}, ISBN \bibinfo{isbn}{9780123744579}.
	
\end{thebibliography}
\end{document}